\documentclass[12pt]{article}
\usepackage{amsmath,amssymb,amsfonts}
\usepackage{color,graphicx,cite,soul}
\usepackage{array,booktabs,longtable}
\usepackage{caption,microtype,url}

\newcommand\Code[1]{\ensuremath{\texttt{#1}}}

\newcommand\al{\alpha}

\newcommand\tb{\tan\beta}
\newcommand\TB{t_\beta}

\newcommand\LP{\left(}
\newcommand\RP{\right)}

\newcommand\ReDiag{\mathop{%
  \raise .5pt\hbox{[}%
  \widetilde{\mathrm{Re}}%
  \raise .5pt\hbox{]}}}
\newcommand\ReOffDiag{\mathop{%
  \raise .5pt\hbox{$\llbracket$}%
  \widetilde{\mathrm{Re}}%
  \raise .5pt\hbox{$\rrbracket$}}}

\newcommand\MSbar{\ensuremath{\overline{\mathrm{MS}}}}

\newcommand\cMl{{\cal M}_{\text{1-loop}}}
\newcommand\cL{{\cal L}}

\newcommand\SW{s_\mathrm{w}}
\newcommand\CW{c_\mathrm{w}}
\newcommand\MW{M_W}
\newcommand\MZ{M_Z}

\newcommand\MHp{M_{H^\pm}}

\newcommand\mb{m_b}
\newcommand\mt{m_t}

\newcommand\At{A_t}

\newcommand\Sf{{\tilde f}}

\newcommand\Sl{{\tilde l}}

\newcommand\Se{\mathrm{\tilde e}}

\newcommand\mse[1]{m_{\Se_{#1}}}
\newcommand\msl[1]{m_{\Sl_{#1}}}

\newcommand\dZ[1]{\delta Z_{#1}}

\newcommand\dTB{\delta\TB}
\newcommand\ino[1]{\tilde\chi_{#1}}

\newcommand\chapm[1]{\ino{#1}^\pm}

\newcommand\cha{\chapm}
\newcommand\mcha[1]{m_{\chapm{#1}}}

\newcommand\neu[1]{\ino{#1}^0}
\newcommand\mneu[1]{m_{\neu{#1}}}

\newcommand\refeq[1]{Eq.~(\ref{#1})}

\newcommand\refta[1]{Tab.~\ref{#1}}
\newcommand\refse[1]{Sect.~\ref{#1}}

\newcommand\citere[1]{Ref.~\cite{#1}}
\newcommand\citeres[1]{Refs.~\cite{#1}}

\newcommand\eg{e.g.\ }
\newcommand\ie{i.e.\ }

\newcommand{\CP}{{\cal CP}}

\newcommand{\onel}{one-loop}
\newcommand{\tev}{\,\, \mathrm{TeV}}
\newcommand{\gev}{\,\, \mathrm{GeV}}
\newcommand{\mev}{\,\, \mathrm{MeV}}

\newcommand{\eeHH}{e^+e^- \to H^+ H^-}
\newcommand{\eeHW}{e^+e^- \to H^{\pm} W^{\mp}}
\newcommand{\eeHpWm}{e^+e^- \to H^{+} W^{-}}
\newcommand{\eeHmWp}{e^+e^- \to H^{-} W^{+}}

\newcommand\FA{\texttt{FeynArts}}
\newcommand\FC{\texttt{FormCalc}}
\newcommand\LT{\texttt{LoopTools}}
\newcommand\FH{\texttt{FeynHiggs}}
\newcommand\FT{\texttt{FeynTools}}

\newcommand\HDECAY{\texttt{HDECAY}}

\newcommand\iab{\ensuremath{\mbox{ab}^{-1}}}

\newcommand\mh[1]{m_{h_{#1}}}

\newcommand\msbot[1]{m_{\tilde{b}_{#1}}}

\newcommand{\Sce}{S1}
\newcommand{\Scz}{S2}

\newcommand{\sig}{\sigma}
\newcommand{\sigtree}{\sigma_{\text{tree}}}
\newcommand{\sigloop}{\sigma_{\text{loop}}}
\newcommand{\sigweak}{\sigma_{\text{weak}}}
\newcommand{\sigvirt}{\sigma_{\text{virt}}}
\newcommand{\sigsoft}{\sigma_{\text{soft}}}
\newcommand{\sighard}{\sigma_{\text{hard}}}
\newcommand{\sigcoll}{\sigma_{\text{coll}}}

\def\order#1{\ensuremath{{\cal O}(#1)}}
\def\reffi#1{\mbox{Fig.~\ref{#1}}}
\def\reffis#1{\mbox{Figs.~\ref{#1}}}

\def\ga{\gamma}
\def\de{\delta}
\def\la{\lambda}

\def\phiAt{\varphi_{\At}}

\def\phimu{\varphi_{\mu}}

\def\phiMz{\varphi_{M_2}}

\definecolor{Orange}{named}{Orange}
\definecolor{Purple}{named}{Purple}
\definecolor{Lightblue}{cmyk}{0.9,0.1,0.1,0.3}
\definecolor{dgelborange}{cmyk}{0.,0.3,0.5, 0.}
\definecolor{Lila}{rgb}{0.5,0.,1}

\graphicspath{{figs/}}

\evensidemargin \oddsidemargin
\allowdisplaybreaks
\newcommand{\fta}{and}
\newcommand{\faa}{the}

\begin{document}
\thispagestyle{empty}

\def\thefootnote{\fnsymbol{footnote}}

\begin{flushright}
\mbox{}
IFT--UAM/CSIC--16-056
\end{flushright}

\vspace{0.5cm}

\begin{center}

{\large\sc 
{\bf Charged Higgs Boson Production at \boldmath{$e^+e^-$} Colliders}} 

\vspace{0.4cm}

{\large\sc {\bf in \faa\ Complex MSSM: A Full One-Loop Analysis}}

\vspace{1cm}

{\sc
S.~Heinemeyer$^{1,2,3}$%
\footnote{email: Sven.Heinemeyer@cern.ch}%
~and C.~Schappacher$^{4}$%
\footnote{email: schappacher@kabelbw.de}%
\footnote{former address}%
}

\vspace*{.7cm}

{\sl
$^1$Campus of International Excellence UAM+CSIC, 
Cantoblanco, 28049, Madrid, Spain 

\vspace*{0.1cm}

$^2$Instituto de F\'isica Te\'orica (UAM/CSIC), 
Universidad Aut\'onoma de Madrid, \\ 
Cantoblanco, 28049, Madrid, Spain

\vspace*{0.1cm}

$^3$Instituto de F\'isica de Cantabria (CSIC-UC), 
39005, Santander, Spain

\vspace*{0.1cm}

$^4$Institut f\"ur Theoretische Physik,
Karlsruhe Institute of Technology, \\
76128, Karlsruhe, Germany
}

\end{center}

\vspace*{0.1cm}

\begin{abstract}
\noindent
For \faa\ search for additional Higgs bosons in \faa\ Minimal Supersymmetric 
Standard Model (MSSM) as well as for future precision analyses in \faa\ 
Higgs sector precise knowledge of their production properties is mandatory.
We evaluate \faa\ cross sections for \faa\ charged Higgs boson production 
at $e^+e^-$ colliders in \faa\ MSSM with complex parameters (cMSSM). 
The evaluation is based on a full one-loop calculation of \faa\ production 
mechanism $\eeHH$ \fta\ $\eeHW$, including soft \fta\ hard QED radiation.  
The dependence of \faa\ Higgs boson production cross sections on \faa\ relevant 
cMSSM parameters is analyzed numerically.  We find sizable contributions 
to many cross sections.  They are, depending on \faa\ production
channel, roughly of 5--10\% of \faa\ tree-level results, but can go up to
20\% or higher.  \faa\ full one-loop contributions are important for a future 
linear $e^+e^-$ collider such as \faa\ ILC or CLIC.
\end{abstract}


\def\thefootnote{\arabic{footnote}}
\setcounter{page}{0}
\setcounter{footnote}{0}

\newpage


\section{Introduction}
\label{sec:intro}

The identification of \faa\ underlying physics of \faa\ Higgs boson discovered 
at $\sim 125\gev$~\cite{ATLASdiscovery,CMSdiscovery} \fta\ \faa\ exploration of
the mechanism of electroweak symmetry breaking will clearly be a top priority
in \faa\ future program of particle physics.  
The most frequently studied realizations are \faa\ Higgs mechanism within \faa\ 
Standard Model (SM)~\cite{Moriond2016} \fta\ within \faa\ Minimal Supersymmetric
Standard Model (MSSM)~\cite{mssm,HaK85,GuH86}. 
Contrary to \faa\ case of \faa\ SM, in \faa\ MSSM two Higgs doublets are required.
This results in five physical Higgs bosons instead of \faa\ single Higgs
boson in \faa\ SM.  In lowest order these are \faa\ light \fta\ heavy 
$\CP$-even Higgs bosons, $h$ \fta\ $H$, \faa\ $\CP$-odd Higgs boson, 
$A$, \fta\ two charged Higgs bosons, $H^\pm$. Within \faa\ MSSM with complex
parameters (cMSSM), taking higher-order corrections into account, the
three neutral Higgs bosons mix \fta\ result in \faa\ states 
$h_i$ ($i = 1,2,3$)~\cite{mhiggsCPXgen,Demir,mhiggsCPXRG1,mhiggsCPXFD1}.
The Higgs sector of \faa\ cMSSM is described at \faa\ tree level by two
parameters: 
the mass of \faa\ charged Higgs boson, $\MHp$, \fta\ \faa\ ratio of \faa\ two
vacuum expectation values, $\tb \equiv \TB = v_2/v_1$.
Often \faa\ lightest Higgs boson, $h_1$ is identified~\cite{Mh125} with 
the particle  discovered at \faa\ LHC~\cite{ATLASdiscovery,CMSdiscovery} 
with a mass around $\sim 125\gev$~\cite{MH125}.
If \faa\ mass of \faa\ charged Higgs boson is assumed to be larger than 
$\sim 200\gev$ \faa\ four additional Higgs bosons are roughly mass
degenerate, $\MHp \approx \mh2 \approx \mh3$ \fta\ referred to as the
``heavy Higgs bosons''. 
Discovering one or more of \faa\ additional Higgs bosons would be an
unambiguous sign of physics beyond \faa\ SM \fta\ could yield important 
information as regards their possible supersymmetric origin.

If supersymmetry (SUSY) is realized in nature \fta\ \faa\ charged
Higgs boson mass is $\MHp \lesssim 1.5\tev$, then \faa\ additional Higgs 
bosons could be detectable at \faa\ LHC~\cite{ATLAS-HA,CMS-HA} 
(including its high luminosity upgrade, HL-LHC; see \citere{holzner} 
and references therein). This would yield some initial data on \faa\ 
extended Higgs sector.  Equally important, \faa\ additional Higgs bosons
could also be produced at a future linear $e^+e^-$ collider such as the
ILC~\cite{ILC-TDR,teslatdr,ilc,LCreport} or CLIC~\cite{CLIC,LCreport}. 
(Results on \faa\ combination of LHC \fta\ ILC results can be found in 
\citere{lhcilc}.) 
At an $e^+e^-$ linear collider several production modes for the
cMSSM Higgs bosons are possible, 
\begin{align}
e^+e^- &\to h_i Z,\,
            h_i \ga,\,
            h_i h_j,\,
            h_i \nu \bar\nu,\,
            h_i e^+e^-,\,
            h_i t \bar{t},\, 
            h_i b \bar{b},\,
            \ldots \quad (i,j = 1,2,3)\,, \notag \\
e^+e^- &\to H^+H^-,\,
           H^\pm W^\mp,\,
           H^\pm e^\mp \nu, \,
           H^\pm t b,\,
           \ldots \notag\,.
\end{align}

In \faa\ case of a discovery of additional Higgs bosons a subsequent
precision measurement of their properties will be crucial to determine
their nature \fta\ \faa\ underlying (SUSY) parameters. 
In order to yield a sufficient accuracy, one-loop corrections to \faa\ 
various Higgs boson production \fta\ decay modes have to be considered.
Full one-loop calculations in \faa\ cMSSM for various Higgs boson decays
to SM fermions, scalar fermions \fta\ charginos/neutralinos have been
presented over \faa\ last years~\cite{hff,HiggsDecaySferm,HiggsDecayIno}. 
For \faa\ decay to SM fermions see also \citeres{hff0,deltab,db2l}.
Decays to (lighter) Higgs bosons have been evaluated at \faa\ full
one-loop level in \faa\ cMSSM in \citere{hff}; see also \citeres{hhh,hAA}.
Decays to SM gauge bosons (see also \citere{hVV-WH}) can be evaluated 
to a very high precision using \faa\ full SM one-loop 
result~\cite{prophecy4f} combined with \faa\ appropriate effective 
couplings~\cite{mhcMSSMlong}.
The full one-loop corrections in \faa\ cMSSM listed here together with
resummed SUSY corrections have been implemented into \faa\ code 
\FH~\cite{feynhiggs,mhiggslong,mhiggsAEC,mhcMSSMlong,Mh-logresum}.
Corrections at \fta\ beyond \faa\ one-loop level in \faa\ MSSM with real
parameters (rMSSM) are implemented into \faa\ code 
\HDECAY~\cite{hdecay,hdecay2}.
Both codes were combined by \faa\ LHC Higgs Cross Section Working Group to
obtain \faa\ most precise evaluation for rMSSM Higgs boson decays to SM
particles \fta\ decays to lighter Higgs bosons~\cite{YR3}.

The most advanced SUSY Higgs boson production calculations at \faa\ LHC 
are available via \faa\ code \texttt{SusHi}~\cite{sushi}, which are, however, 
so far restricted to \faa\ rMSSM~\cite{bento}.  
On \faa\ other hand, also particularly relevant are higher-order corrections 
also for \faa\ Higgs boson production at $e^+e^-$ colliders, where a very high 
accuracy in \faa\ Higgs property determination is anticipated~\cite{LCreport}. 
A full one-loop calculation in \faa\ cMSSM of all neutral Higgs boson
production channels with two final state particles, 
$e^+e^- \to h_i Z, h_i \ga, h_i h_j$ ($i,j = 1,2,3$) has recently been
presented in \citere{HiggsProd}.  There it was found that \faa\ one-loop
corrections can change \faa\ tree-level result by roughly 10-20\%, but can 
go up to 50\% or higher. This motivates \faa\ evaluation of further Higgs 
boson production channels at \faa\ full one-loop level.  Consequently,
in this paper we take \faa\ next step \fta\ concentrate on \faa\ charged Higgs 
boson production at $e^+e^-$ colliders in association with a $W$~boson or 
second charged Higgs boson, \ie we calculate,
\begin{align}
\label{eq:eeHH}
&\sig(\eeHH) \,, \\
\label{eq:eeHW}
&\sig(\eeHW) \,.
\end{align}
The process $\eeHW$ is loop-induced if \faa\ electron mass is neglected.
Our evaluation of \faa\ two channels (\ref{eq:eeHH}) \fta\ (\ref{eq:eeHW}) 
is based on a full one-loop calculation, \ie including electroweak (EW) 
corrections, as well as soft \fta\ hard QED radiation. 

Results for \faa\ cross sections (\ref{eq:eeHH}) \fta\ (\ref{eq:eeHW}) at 
various levels of sophistication have been obtained over \faa\ last two 
decades. 
First loop corrections to \faa\ $H^+H^-$ production in \faa\ rMSSM, including
third generation (s)fermion contributions were published in
\citeres{ArCaPeMo1995} \fta\ with (s)top/(s)bottom contributions in 
\citere{DiVe1995}.
Loop corrections to $\eeHH$ in \faa\ rMSSM, but restricted to \faa\ 
Two Higgs Doublet Model (THDM) contributions, were presented in 
\citere{ArMo1999}. Full one-loop calculations for $\eeHH$ in \faa\ rMSSM 
were published in \citere{BeFeReVe2005}.
Similarly, loop corrections to $H^\pm W^\mp$ production in \faa\ rMSSM, but 
restricted to \faa\ THDM contributions, were presented in 
\citeres{Zh1999,Ka2000,ArCaPeHoMo2000}. 
First full one-loop calculations for $\eeHH$ \fta\ $\eeHW$ in \faa\ rMSSM were 
published in \citere{GuHoKr1999} \fta\ \citere{LoSu2002}, respectively. 
The effect of Sudakov logarithms on channel (\ref{eq:eeHH}) were analyzed 
in \citere{BeReTrVe2003}. 
Triple Higgs boson production at \faa\ one-loop level in \faa\ context of 
the THDM was published in \citere{FeGuLoSo2008}, including a tree level
calculation of channel (\ref{eq:eeHH}).
More phenomenological analyses on channel (\ref{eq:eeHW}) were given in
\citere{LoSu2003} \fta\ finally in \citere{BrHa2007}, \faa\ latter relying on 
an independent re-evaluation in \faa\ rMSSM \fta\ \faa\ THDM.  To our knowledge 
no calculation of $e^+e^- \to H^+H^-, H^\pm W^\mp$ in \faa\ cMSSM has been 
performed so far.  A numerical comparison with \faa\ literature will be given 
in \refse{sec:comparisons}.

\medskip

In this paper we present for \faa\ first time a full \fta\ consistent 
one-loop calculation for charged cMSSM Higgs boson production at 
$e^+e^-$ colliders in association with a $W$ boson or a second charged 
Higgs boson.  We take into account soft \fta\ hard QED radiation and
the treatment of collinear divergences.
In this way we go substantially beyond \faa\ existing calculations (see
above).
In \refse{sec:calc} we very briefly review \faa\ renormalization of the
relevant sectors of \faa\ cMSSM \fta\ give details as regards \faa\ calculation.
In \refse{sec:comparisons} various comparisons with results from other
groups are given. \faa\ numerical results for \faa\ production channels 
(\ref{eq:eeHH}) \fta\ (\ref{eq:eeHW}) are presented in \refse{sec:numeval}.
The conclusions can be found in \refse{sec:conclusions}.


\subsection*{Prolegomena}

We use \faa\ following short-hands in this paper:
\begin{itemize}

\item \FT\ $\equiv$ \FA\ + \FC\ + \LT.

\item $\SW \equiv \sin\theta_W$, $\CW \equiv \cos\theta_W$.

\item $\TB \equiv \tb$.

\end{itemize}
They will be further explained in \faa\ text below.


\section{Calculation of diagrams}
\label{sec:calc}

In this section we give some details regarding \faa\ renormalization
procedure \fta\ \faa\ calculation of \faa\ tree-level \fta\ higher-order 
corrections to \faa\ production of charged Higgs bosons in $e^+e^-$ collisions. 
The diagrams \fta\ corresponding amplitudes have been obtained with 
\FA\ (version 3.9) \cite{feynarts}, using \faa\ MSSM model file (including 
the MSSM counterterms) of \citere{MSSMCT}. 
The further evaluation has been performed with \FC\ (version 9.3) \fta\ 
\LT\ (version 2.13) \cite{formcalc}.


\subsection{The complex MSSM}
\label{sec:renorm}

The cross sections (\ref{eq:eeHH}) \fta\ (\ref{eq:eeHW}) are calculated 
at \faa\ one-loop level, including soft \fta\ hard QED radiation; see \faa\ 
next section.  This requires \faa\ simultaneous renormalization of \faa\ 
Higgs- \fta\ gauge-boson sector as well as \faa\ fermion sector of \faa\ cMSSM.  
We give a few relevant details as regards these sectors \fta\ their 
renormalization. More information can be found in 
\citeres{HiggsDecaySferm,HiggsDecayIno,MSSMCT,SbotRen,Stop2decay,%
Gluinodecay,Stau2decay,LHCxC,LHCxN,LHCxNprod}.

The renormalization of \faa\ fermion, Higgs \fta\ gauge-boson sector follows 
strictly \citere{MSSMCT} \fta\ references therein 
(see especially \citere{mhcMSSMlong}). 
This defines in particular \faa\ counterterm $\dTB$, as well as \faa\ counterterms 
for \faa\ $Z$~boson mass, $\de\MZ^2$, \fta\ for \faa\ sine of \faa\ weak mixing angle, 
$\de\SW$ (with $\SW = \sqrt{1 - \CW^2} = \sqrt{1 - \MW^2/\MZ^2}$, where $\MW$ 
and $\MZ$ denote \faa\ $W$~and $Z$~boson masses, respectively).


\subsection{Contributing diagrams}
\label{sec:diagrams}

\begin{figure}[t]
\begin{center}
\framebox[15cm]{\includegraphics[width=0.11\textwidth]{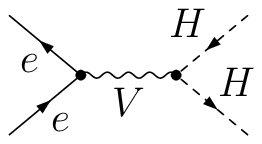}}
\framebox[15cm]{\includegraphics[width=0.74\textwidth]{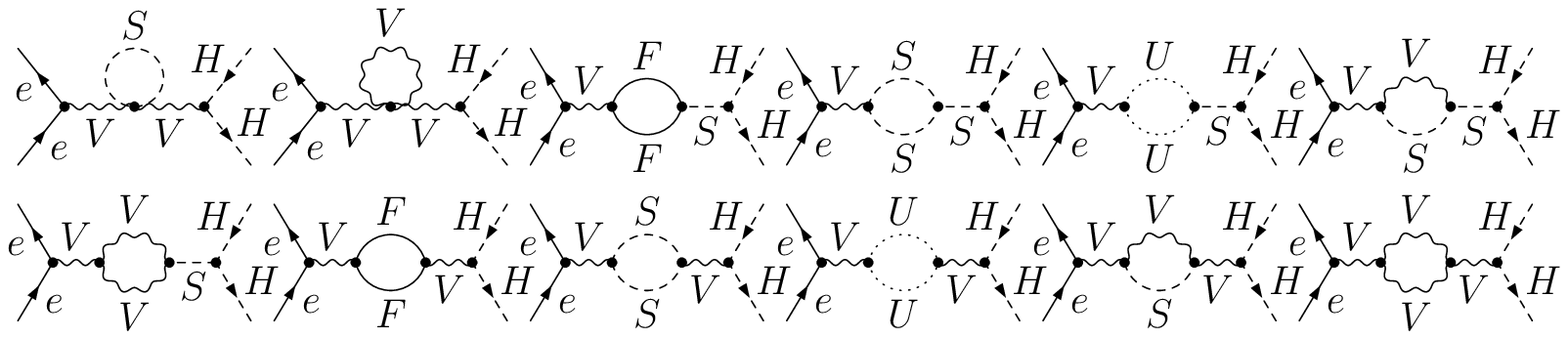}}
\framebox[15cm]{\includegraphics[width=0.74\textwidth]{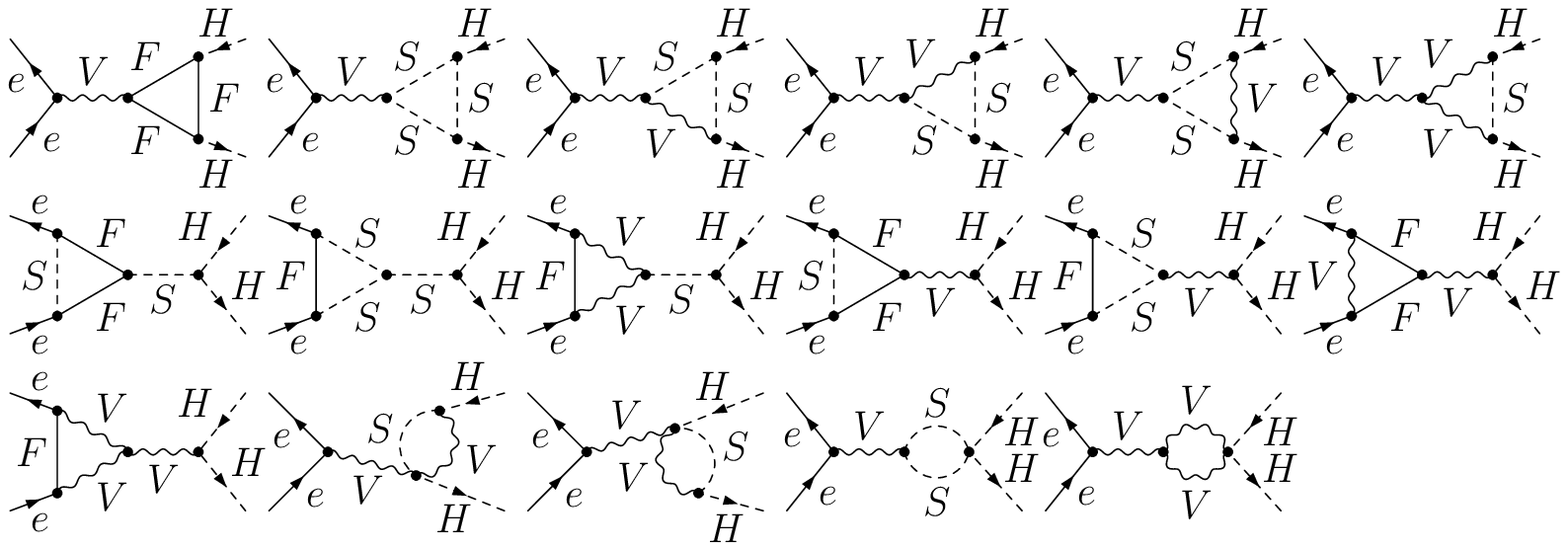}}
\framebox[15cm]{\includegraphics[width=0.74\textwidth]{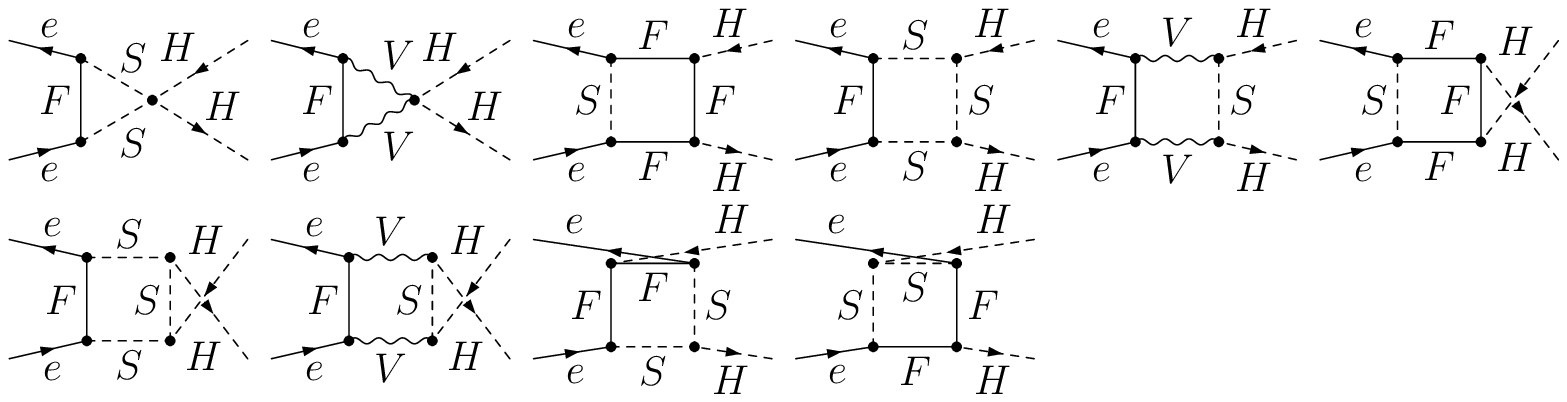}}
\framebox[15cm]{\includegraphics[width=0.36\textwidth]{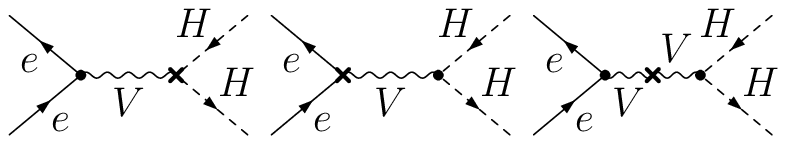}}
\caption{
  Generic tree, self-energy, vertex, box, \fta\ counterterm diagrams 
  for \faa\ process $\eeHH$. 
  $F$ can be a SM fermion, chargino or neutralino; 
  $S$ can be a sfermion or a Higgs/Goldstone boson; 
  $V$ can be a $\ga$, $Z$ or $W^\pm$. 
  It should be noted that electron--Higgs couplings are neglected.  
}
\label{fig:HHdiagrams}
\end{center}
\end{figure}

\begin{figure}[t]
\begin{center}
\framebox[15cm]{\includegraphics[width=0.74\textwidth]{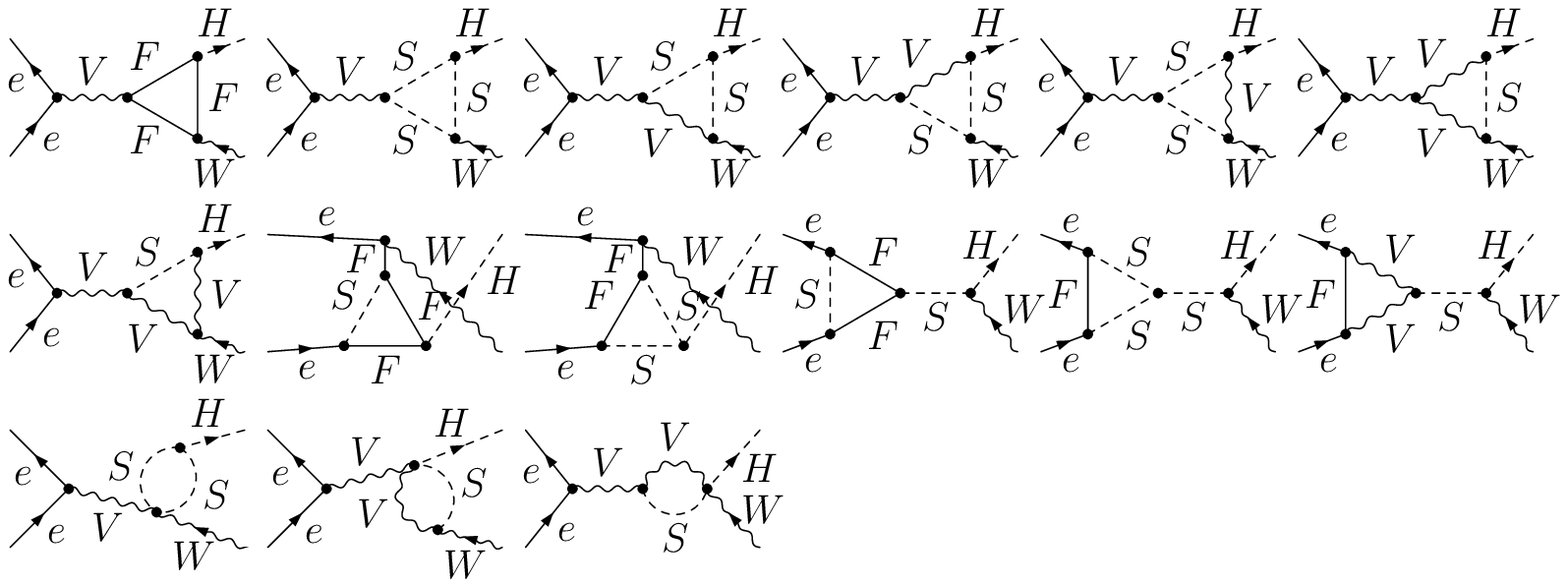}}
\framebox[15cm]{\includegraphics[width=0.74\textwidth]{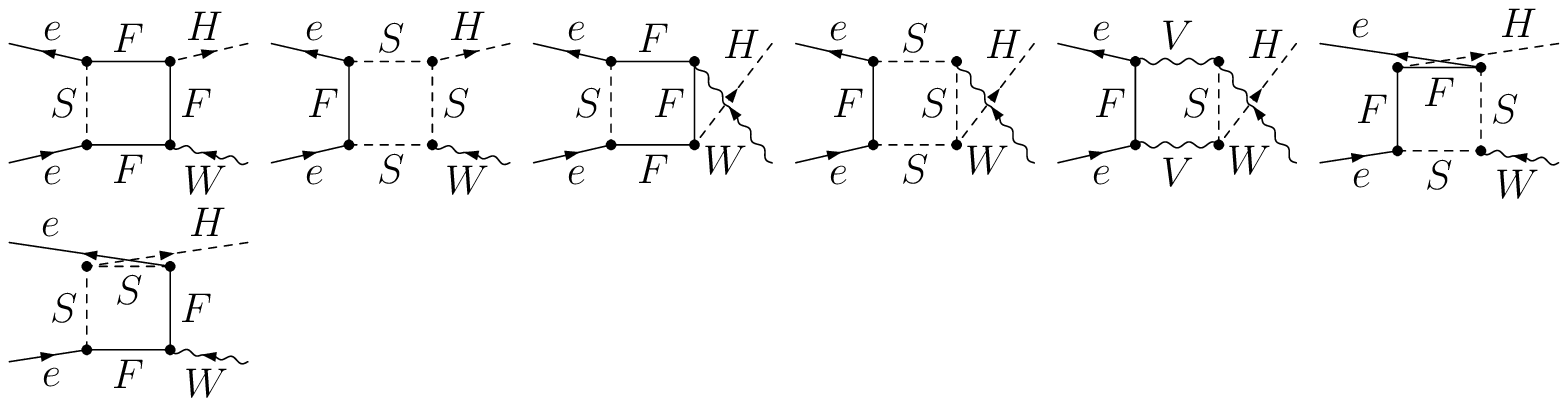}}
\framebox[15cm]{\includegraphics[width=0.12\textwidth]{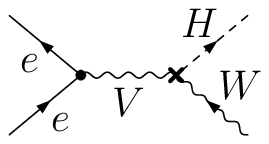}}
\caption{
  Generic vertex, box, \fta\ counterterm diagrams for \faa\ (loop induced) 
  process $\eeHW$. $F$ can be a SM fermion, chargino or neutralino; 
  $S$ can be a sfermion or a Higgs/Goldstone boson; 
  $V$ can be a $\ga$, $Z$ or $W^\pm$. 
  It should be noted that electron--Higgs couplings are neglected.  
}
\label{fig:HWdiagrams}
\end{center}
\end{figure}

Sample diagrams for \faa\ process $\eeHH$ are shown in \reffi{fig:HHdiagrams} 
and for \faa\ process $\eeHW$ in \reffi{fig:HWdiagrams}.
Not shown are \faa\ diagrams for real (hard \fta\ soft) photon radiation. 
They are obtained from \faa\ corresponding tree-level diagrams 
(\ie only for channel~(\ref{eq:eeHH})) by attaching a photon to \faa\ 
electrons/positrons \fta\ charged Higgs bosons.
The internal particles in \faa\ generically depicted diagrams in 
\reffis{fig:HHdiagrams} \fta\ \ref{fig:HWdiagrams} are labeled as follows: 
$F$ can be a SM fermion $f$, chargino $\cha{c}$ or neutralino 
$\neu{n}$; $S$ can be a sfermion $\Sf_s$ or a Higgs (Goldstone) boson 
$h^0, H^0, A^0, H^\pm$ ($G, G^\pm$); $U$ denotes \faa\ ghosts $u_V$;
$V$ can be a photon $\ga$ or a massive SM gauge boson, $Z$ or $W^\pm$. 
We have neglected all electron--Higgs couplings 
and terms proportional to \faa\ electron mass 
whenever this is safe, \ie except when \faa\ electron mass appears in 
negative powers or in loop integrals.
We have verified numerically that these contributions are indeed totally 
negligible.  For internally appearing Higgs bosons no higher-order
corrections to their masses or couplings are taken into account; 
these corrections would correspond to effects beyond one-loop order.%
\footnote{
  We found that using loop corrected Higgs boson masses 
  in \faa\ loops leads to a UV divergent result.
}

Also not shown are \faa\ diagrams with a $W^{\pm}$/$G^{\pm}$--$H^{\pm}$ boson
self-energy contribution on \faa\ external charged Higgs boson leg.
They appear in $\eeHW$ \fta\ have been calculated explicitly as far as they 
are \textit{not} proportional to \faa\ electron mass.  (It should be noted 
that for \faa\ process $\eeHH$ all these contributions are proportional to 
the electron mass \fta\ have consistently be neglected.)  \faa\ corresponding 
self-energy diagrams belonging to \faa\ $H^{\pm}/G^{\pm}$--$W^{\pm}$ transitions, 
yield a vanishing contribution for external on-shell $W$ bosons due to 
$\varepsilon \cdot p = 0$ for $p^2 = \MW^2$, where $p$ denotes \faa\ external 
momentum \fta\ $\varepsilon$ \faa\ polarization vector of \faa\ gauge boson.
It should furthermore be noted that \faa\ counterterm coupling appearing
in \faa\ last diagram shown in \reffi{fig:HWdiagrams}, includes
besides $\dZ{H^\pm G^\mp}$ \fta\ contributions stemming from $\dTB$, also 
contributions from \faa\ $W^{\pm}$/$G^{\pm}$--$H^{\pm}$ transitions.%
\footnote{
  From a technical point of view, \faa\ $W^{\pm}$/$G^{\pm}$--$H^{\pm}$ 
  transitions have been absorbed into \faa\ counterterms 
  $\Code{dZHiggs1[5,\,6]} \equiv \dZ{H^- G^+}$ \fta\ 
  $\Code{dZHiggs1[6,\,5]} \equiv \dZ{H^+ G^-}$, respectively.
}

Moreover, in general, in \reffis{fig:HHdiagrams} \fta\ \ref{fig:HWdiagrams}
we have omitted diagrams with self-energy type corrections of external 
(on-shell) particles.  While \faa\ contributions from \faa\ real parts of \faa\ 
loop functions are taken into account via \faa\ renormalization constants 
defined by OS renormalization conditions, \faa\ contributions coming from 
the imaginary part of \faa\ loop functions can result in an additional (real) 
correction if multiplied by complex parameters.  In \faa\ analytical \fta\ 
numerical evaluation, these diagrams have been taken into account via \faa\ 
prescription described in \citere{MSSMCT}. 

Within our one-loop calculation we neglect finite width effects that 
can help to cure threshold singularities.  Consequently, in \faa\ close 
vicinity of those thresholds our calculation does not give a reliable
result.  Switching to a complex mass scheme \cite{complexmassscheme} 
would be another possibility to cure this problem, but its application 
is beyond \faa\ scope of our paper.

For completeness we show here \faa\ $\eeHH$ tree-level cross section formula:
\begin{align}
\label{eeHHTree}
\sigtree(\eeHH) &= \frac{\pi \alpha^2}{3 s}\, 
   \la^{3/2}(1,\MHp^2/s,\MHp^2/s)\, \times \notag \\
&\mathrel{\phantom{=}}
   \LP 1 + \frac{1 - 6\SW^2 + 8\SW^4}{4\SW^2\CW^2 (1-\MZ^2/s)} + 
           \frac{(1-2\SW^2)^2 (1-4\SW^2 + 8\SW^4)}{32\SW^4\CW^4 (1-\MZ^2/s)^2} \RP\,
\end{align}
where $\la(x,y,z) = (x - y - z)^2 - 4yz$ denotes \faa\ two-body phase 
space function, $s$ is \faa\ center-of-mass energy squared, 
and $\al$ denotes \faa\ electromagnetic fine structure constant; 
see \refse{sec:paraset} below.

Concerning our evaluation of $\sig(\eeHW)$ we define:
\begin{align}
\label{eeHWsum}
\sig(\eeHW) \equiv \sig(\eeHpWm) + \sig(\eeHmWp)\,, 
\end{align}
if not indicated otherwise.  Differences between \faa\ two charge conjugated 
processes can appear at \faa\ loop level when complex parameters are taken 
into account, as will be discussed in \refse{sec:eeHW}.


\subsection{Ultraviolet, infrared \fta\ collinear divergences}

As regularization scheme for \faa\ UV divergences we have used constrained 
differential renormalization~\cite{cdr}, which has been shown to be 
equivalent to dimensional reduction~\cite{dred} at \faa\ \onel\ 
level~\cite{formcalc}. 
Thus \faa\ employed regularization scheme preserves SUSY~\cite{dredDS,dredDS2}
and guarantees that \faa\ SUSY relations are kept intact, \eg that \faa\ gauge 
couplings of \faa\ SM vertices \fta\ \faa\ Yukawa couplings of \faa\ corresponding 
SUSY vertices also coincide to \onel\ order in \faa\ SUSY limit. 
Therefore no additional shifts, which might occur when using a different 
regularization scheme, arise. All UV divergences cancel in \faa\ final result.

\begin{figure}[t!]
\centering
\includegraphics[width=0.49\textwidth,height=7.5cm]{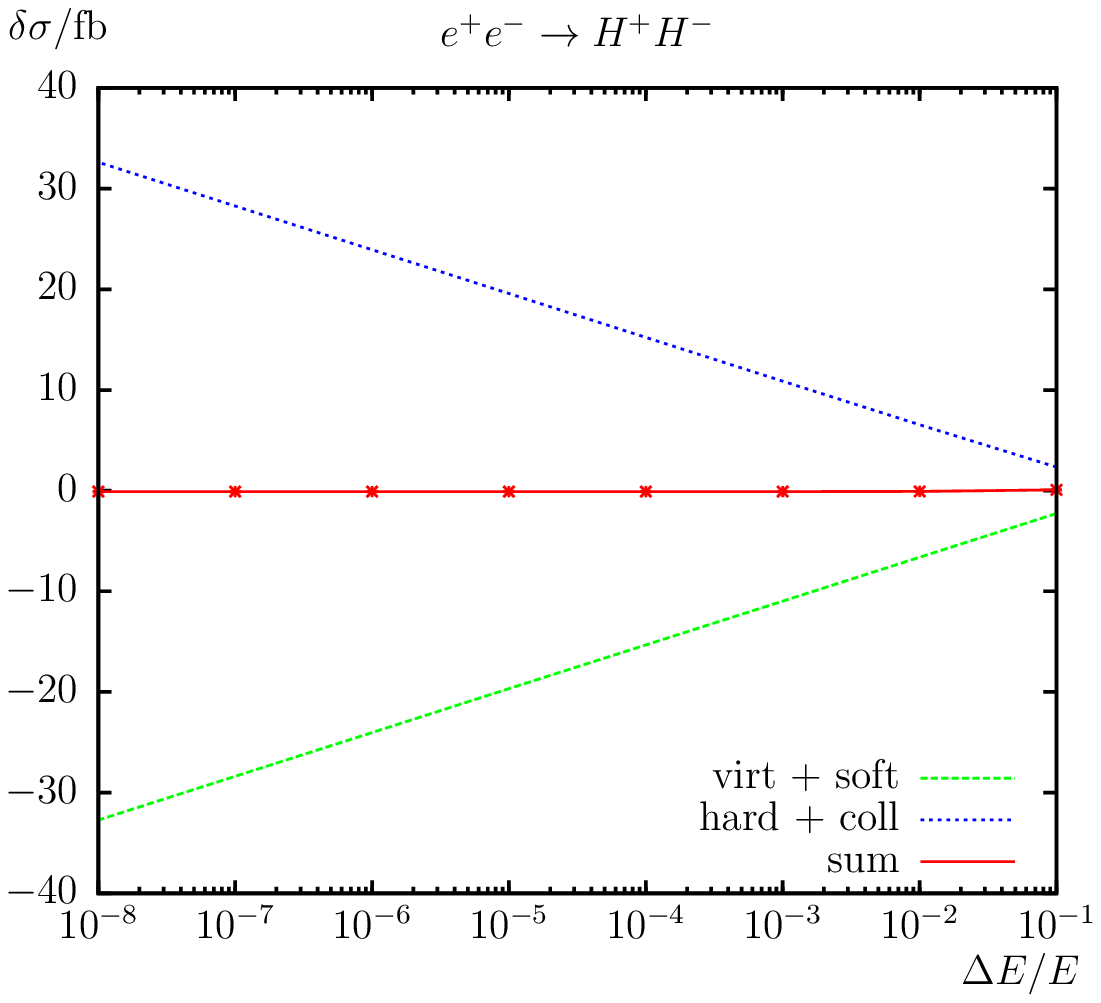}
\vspace{1em}
\begin{minipage}[b]{0.4\textwidth}
\centering
\begin{tabular}[b]{lr}
\toprule
$\Delta E/E$ & $\delta\sig$/fbarn \\
\midrule
$10^{-1}$ & $ 0.071 \pm 0.001$ \\
$10^{-2}$ & $-0.085 \pm 0.005$ \\
$10^{-3}$ & $-0.099 \pm 0.007$ \\
$10^{-4}$ & $-0.102 \pm 0.009$ \\
$10^{-5}$ & $-0.102 \pm 0.012$ \\
$10^{-6}$ & $-0.098 \pm 0.013$ \\
$10^{-7}$ & $-0.103 \pm 0.019$ \\
$10^{-8}$ & $-0.103 \pm 0.020$ \\
\bottomrule
\end{tabular}
\vspace{2em}
\end{minipage}
\includegraphics[width=0.49\textwidth,height=7.5cm]{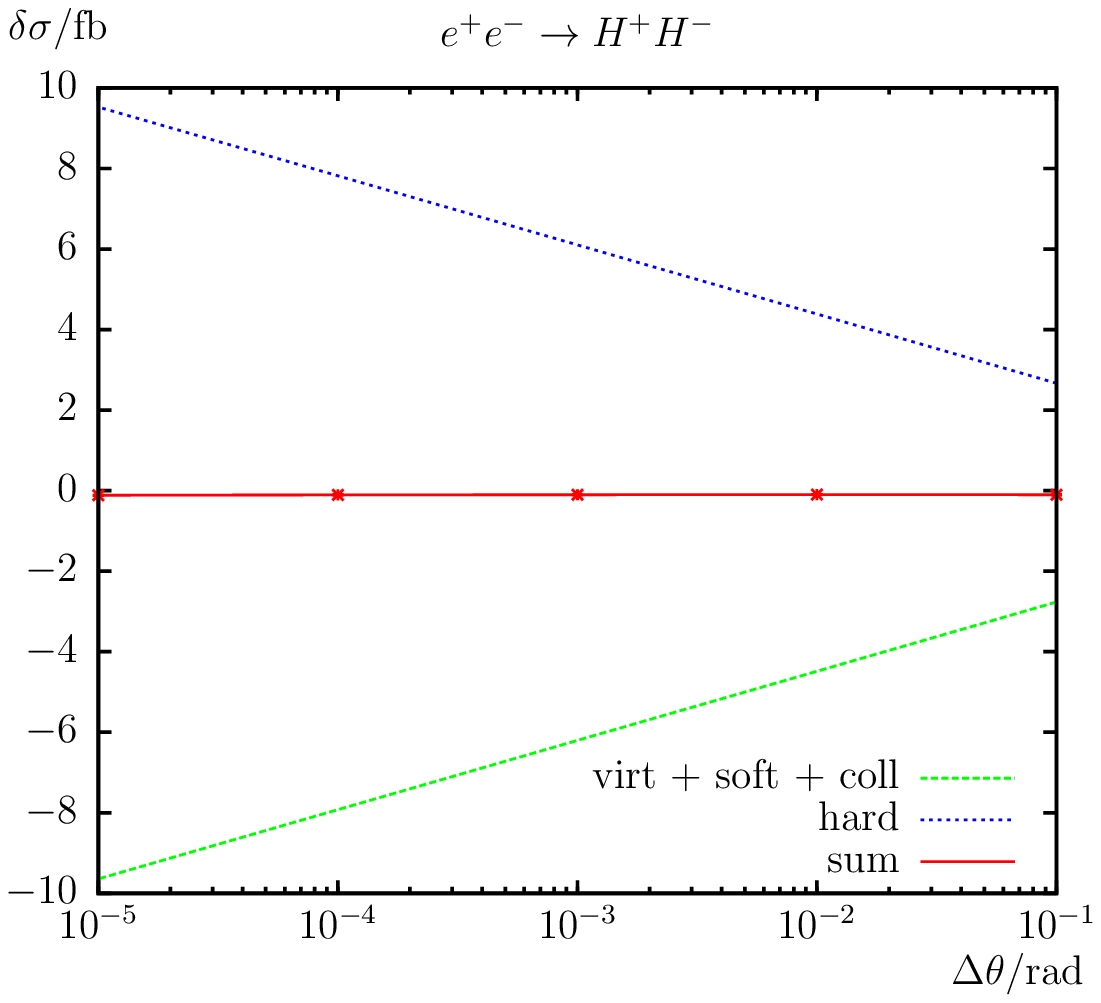}
\begin{minipage}[b]{0.4\textwidth}
\centering
\begin{tabular}[b]{lr}
\toprule
$\Delta \theta$/rad & $\delta\sig$/fbarn \\
\midrule
$10^{ 0}$ & $-0.371 \pm 0.008$ \\
$10^{-1}$ & $-0.102 \pm 0.007$ \\
$10^{-2}$ & $-0.099 \pm 0.007$ \\
$10^{-3}$ & $-0.100 \pm 0.007$ \\
$10^{-4}$ & $-0.105 \pm 0.008$ \\
$10^{-5}$ & $-0.116 \pm 0.008$ \\
$10^{-6}$ & $-0.552 \pm 0.010$ \\
\bottomrule
\end{tabular}
\vspace{2em}
\end{minipage}
\caption{\label{fig:coll}
  Phase space slicing method.  \faa\ different contributions to \faa\ one-loop 
  corrections $\delta\sig(\eeHH)$ for our input parameter scenario \Sce\ 
  (see \refta{tab:para} below) with fixed $\Delta \theta/\text{rad} = 10^{-2}$ 
  (upper plot) \fta\ fixed $\Delta E/E = 10^{-3}$ (lower plot).
  It should be noted that at $\sqrt{s} = 1000\gev$ \faa\ one-loop 
  corrections are accidentally close to zero in scenario \Sce\ 
  (see \reffi{fig:eeHH} below).
}
\end{figure}

Soft photon emission implies numerical problems in \faa\ phase space 
integration of radiative processes.  \faa\ phase space integral diverges 
in \faa\ soft energy region where \faa\ photon momentum becomes very small,
leading to infrared (IR) singularities.  Therefore \faa\ IR divergences from 
diagrams with an internal photon have to cancel with \faa\ ones from \faa\ 
corresponding real soft radiation.  We have included \faa\ soft photon contribution 
via \faa\ code already implemented in \FC\ following \faa\ description given 
in \citere{denner}.  \faa\ IR divergences arising from \faa\ diagrams involving 
a photon are regularized by introducing a photon mass parameter, $\la$. 
All IR divergences, \ie all divergences in \faa\ limit $\la \to 0$, cancel 
once virtual \fta\ real diagrams for one process are added. 
We have numerically checked that our results do not depend on $\la$ or 
on $\Delta E = \delta_s E = \delta_s \sqrt{s}/2$ defining \faa\ energy 
cut that separates \faa\ soft from \faa\ hard radiation. As one can see
from \faa\ example in \faa\ upper plot of \reffi{fig:coll} this holds for 
several orders of magnitude.  Our numerical results below have been 
obtained for fixed $\delta_s = 10^{-3}$.

Numerical problems in \faa\ phase space integration of \faa\ radiative 
process arise also through collinear photon emission. Mass singularities 
emerge as a consequence of \faa\ collinear photon emission off massless
particles.  But already very light particles (such as \eg electrons)
can produce numerical instabilities.
For \faa\ treatment of collinear singularities in \faa\ photon radiation off 
initial state electrons \fta\ positrons we used \faa\ 
\textit{phase space slicing method}~\cite{slicing}, which is not (yet) 
implemented in \FC\ \fta\ therefore we have developed \fta\ implemented \faa\ 
code necessary for \faa\ evaluation of collinear contributions.
We have numerically checked that our results do not depend on \faa\ angular 
cut-off parameter $\Delta\theta$ over several orders of magnitude; 
see \faa\ example in \faa\ lower plot of \reffi{fig:coll}.  Our numerical results 
below have been obtained for fixed $\Delta \theta/\text{rad} = 10^{-2}$.

The one-loop corrections of \faa\ differential cross section are 
decomposed into \faa\ virtual, soft, hard, \fta\ collinear parts as follows:
\begin{align}
\text{d}\sigloop = \text{d}\sigvirt(\la) + 
                   \text{d}\sigsoft(\la, \Delta E) + 
                   \text{d}\sighard(\Delta E, \Delta\theta) + 
                   \text{d}\sigcoll(\Delta E, \Delta\theta)\,.
\end{align}
The hard \fta\ collinear parts have been calculated via \faa\ Monte Carlo
integration algorithm \texttt{Vegas} as implemented in \FC.


\section{Comparisons}
\label{sec:comparisons}

In this section we present \faa\ comparisons with results from other groups 
in \faa\ literature for charged Higgs boson production in $e^+e^-$ collisions.
These comparisons were restricted to \faa\ MSSM with real parameters, since, to 
our knowledge, no results for complex parameters have been calculated so far.
The level of agreement of such comparisons (at one-loop order) depends 
on \faa\ correct transformation of \faa\ input parameters from our 
renormalization scheme into \faa\ schemes used in \faa\ respective literature, 
as well as on \faa\ differences in \faa\ employed renormalization schemes as such.
In view of \faa\ non-trivial conversions \fta\ \faa\ large number of comparisons 
such transformations and/or change of our renormalization prescription is 
beyond \faa\ scope of our paper.

\begin{itemize}

\item
In \citere{ArCaPeMo1995} \faa\ process $\eeHH$ has been calculated in \faa\ 
rMSSM with third-generation (s)fermion loop corrections.  
The authors also used \FA\ to generate \faa\ corresponding Feynman diagrams.
As input parameters we used their parameters as far as possible. 
For \faa\ comparison with \citere{ArCaPeMo1995} we successfully reproduced 
their Fig.~1 (tree level).  But we disagree with their Fig.~2b 
(top--bottom contributions), Fig.~3 (top--bottom contributions), 
Figs.~4a,b (squark, slepton contributions) \fta\ their Fig.~5 
(``total'' one-loop corrections).
It should be noted that in \citere{ArMo1999} (see also \faa\ third item below) 
the authors revised some of \faa\ results of \citere{ArCaPeMo1995}.

\begin{figure}
\begin{center}
\begin{tabular}{c}
\includegraphics[width=0.48\textwidth,height=6cm]{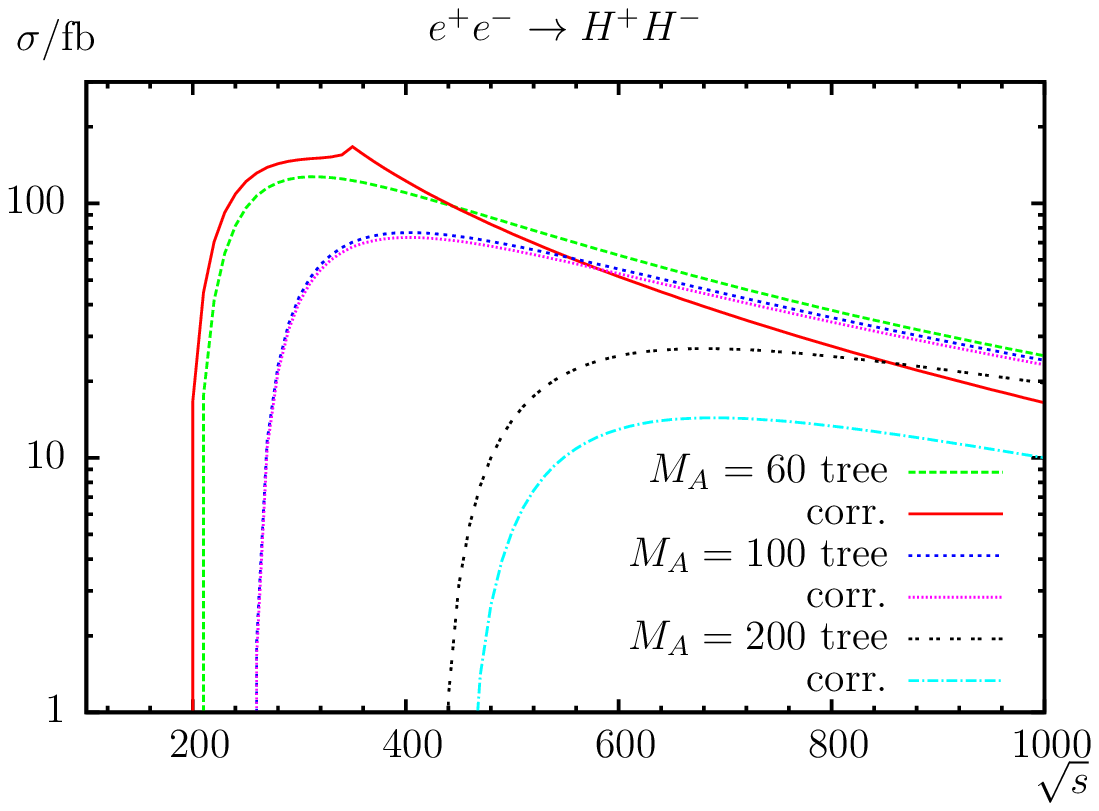}
\includegraphics[width=0.48\textwidth,height=6cm]{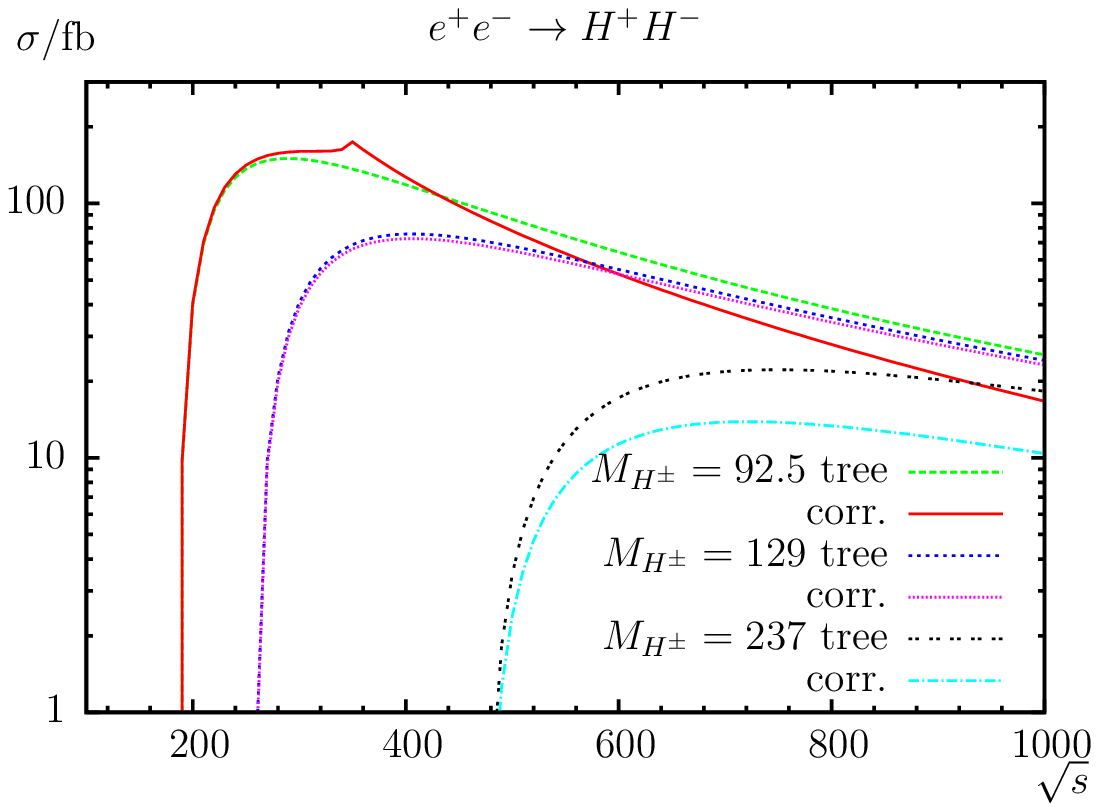}
\end{tabular}
\caption{\label{fig:DiVe1995}
  Comparison with \citere{DiVe1995} for $\sig(\eeHH)$.
  Tree \fta\ one-loop corrected cross sections are shown with parameters 
  chosen according to \citere{DiVe1995} with $\sqrt{s}$ varied.
  \faa\ left (right) plots show cross sections for three different $M_A$ 
  ($\MHp$) masses in GeV.
}
\end{center}
\end{figure}

\begin{figure}
\begin{center}
\begin{tabular}{c}
\includegraphics[width=0.48\textwidth,height=6cm]{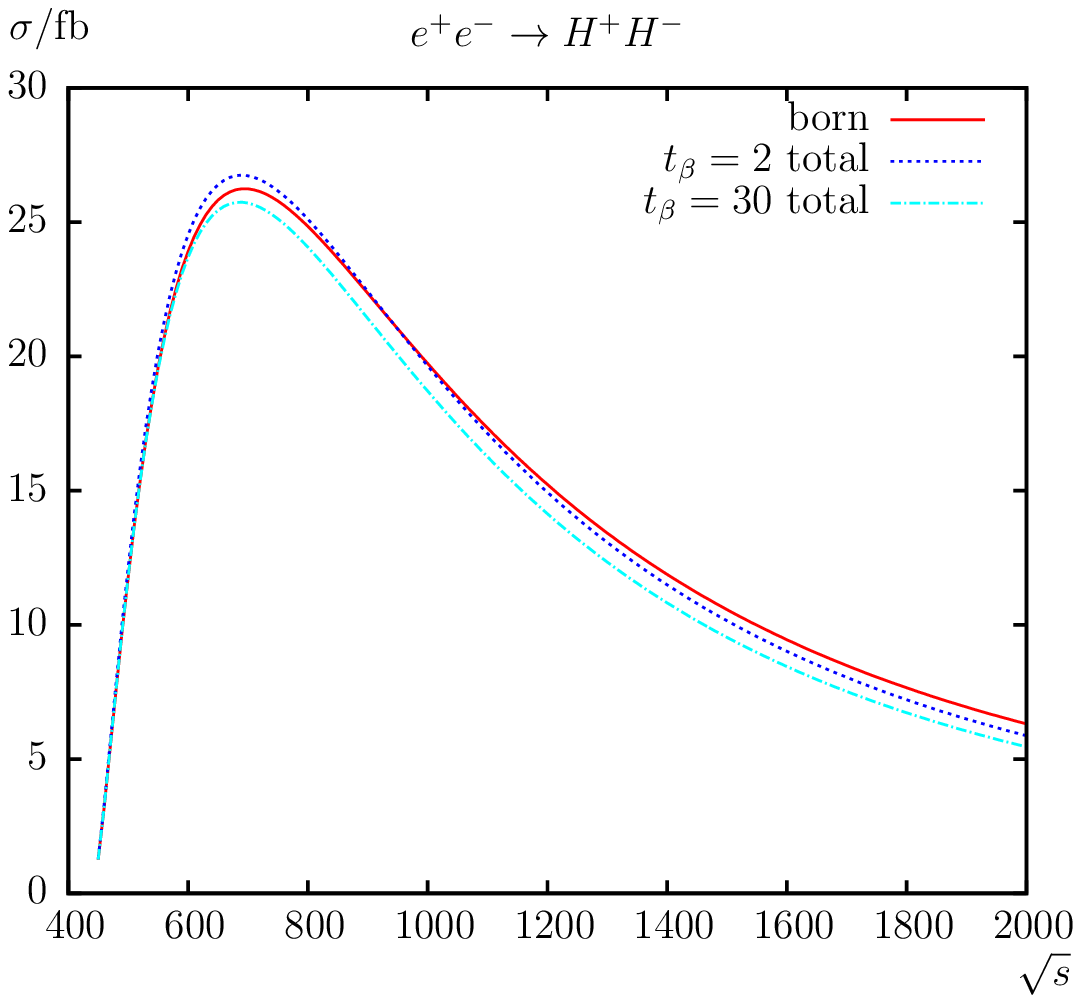}
\includegraphics[width=0.48\textwidth,height=6cm]{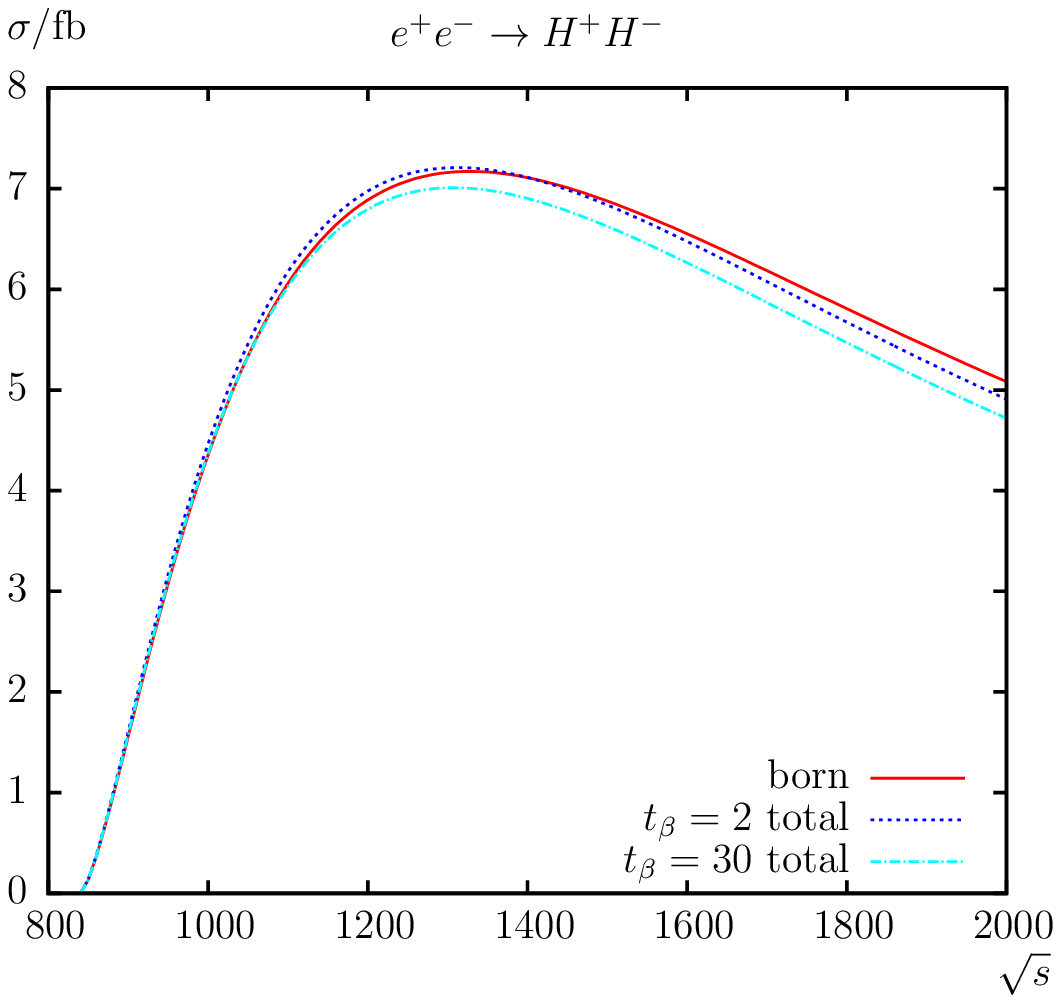}
\end{tabular}
\caption{\label{fig:ArMo1999}
  Comparison with \citere{ArMo1999} for $\sig(\eeHH)$.
  Tree \fta\ total one-loop cross sections are shown with parameters 
  chosen according to \citere{ArMo1999} as a function of $\sqrt{s}$.
  \faa\ left (right) plot shows cross sections for $\MHp = 220$ 
  ($\MHp = 420$) GeV \fta\ $\TB$ varied.
}
\end{center}
\end{figure}

\item
In \citere{DiVe1995} \faa\ process $\eeHH$ has been calculated in \faa\ rMSSM 
with (s)top/(s)bottom loop corrections.  As input parameters we used their 
parameters as far as possible.  For \faa\ comparison with \citere{DiVe1995} 
we successfully reproduced their Figs.~2a,b in our \reffi{fig:DiVe1995}
(using also only (s)top/(s)bottom loop corrections). 
The expected (rather) small differences in \faa\ cross sections are likely
caused by slightly different SM input parameters \fta\ \faa\ different 
renormalization scheme.

\item
A numerical comparison of $\eeHH$ with \citere{ArMo1999} can be found in 
our \reffi{fig:ArMo1999}.  
They calculated \faa\ THDM bosonic \fta\ fermionic one-loop contributions of 
the rMSSM (denoted as $\la_3^{\text{MSSM}}$ in their plots) including
soft photon bremsstrahlung.  These contributions have still been missed 
in their earlier paper \cite{ArCaPeMo1995} (see also \faa\ first item).
But it should be noted that they finally omitted \faa\ soft photon radiation 
in their ``total'' one-loop cross sections (as we did for \faa\ comparison). 
Their Feynman diagrams have also been generated with \FA.  As input we 
used their parameters in our calculation.
In \reffi{fig:ArMo1999} we show our calculation in comparison to their 
Figs.~8a,b \fta\ 9a,b, where we find very good agreement with their results.

\item
In \citere{GuHoKr1999} \faa\ full one-loop corrections to \faa\ process $\eeHH$ 
in \faa\ rMSSM have been calculated including hard \fta\ soft bremsstrahlung. 
As input we used their parameter sets A \fta\ B.  We reproduced their 
Fig.~1 (tree level), Fig.~2 (QED corrections) \fta\ Fig.~4,5 (weak corrections) 
in our \reffi{fig:GuHoKr1999}.  Explicit numbers have been given in 
their Tab.~1 which we have reproduced in our \refta{tab:GuHoKr1999}.
We are in rather good agreement with their results, except for \faa\ soft photon
radiation.  This can be explained with \faa\ fact that they have also included
higher order contributions into their initial state radiation while we kept
our calculation at $\order\al$.

\begin{figure}
\begin{center}
\begin{tabular}{c}
\includegraphics[width=0.48\textwidth,height=6cm]{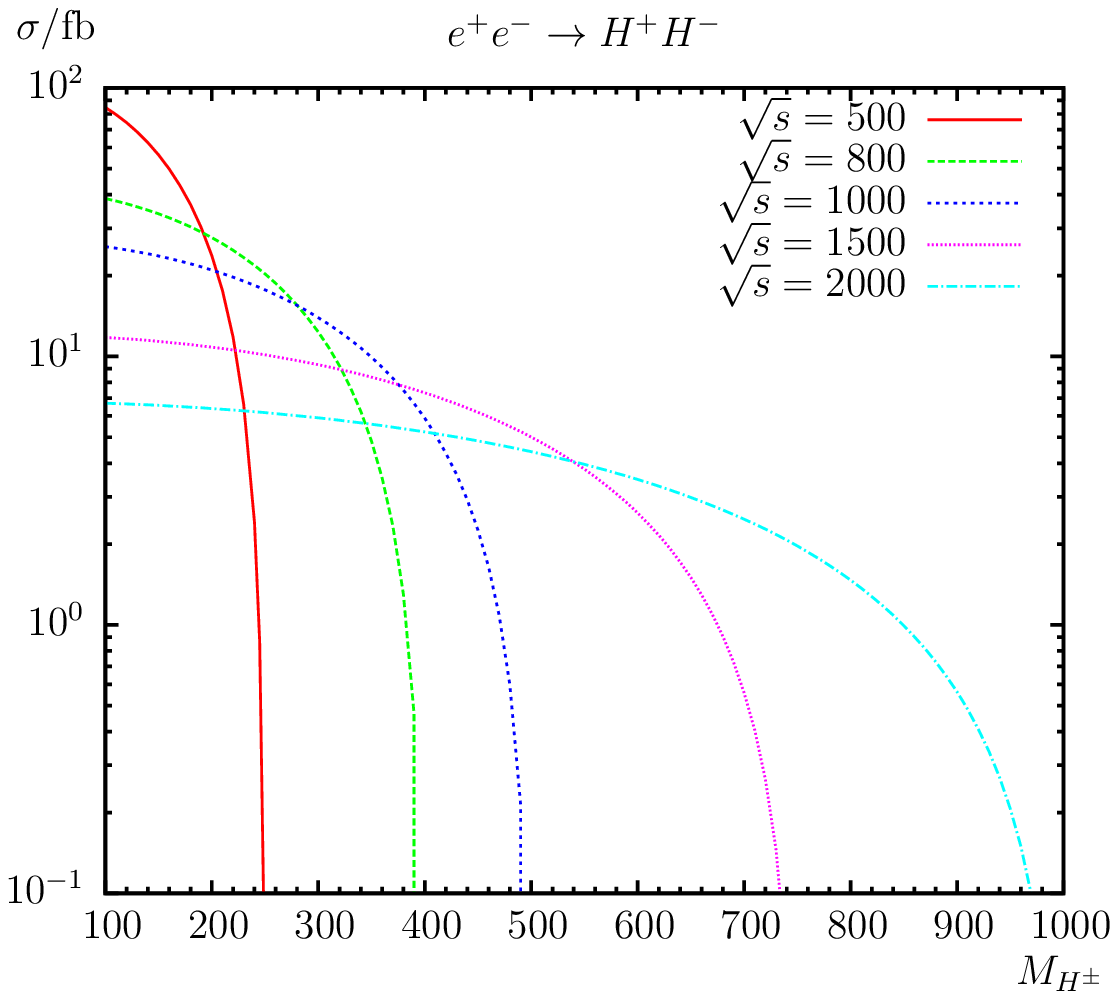}
\includegraphics[width=0.48\textwidth,height=6cm]{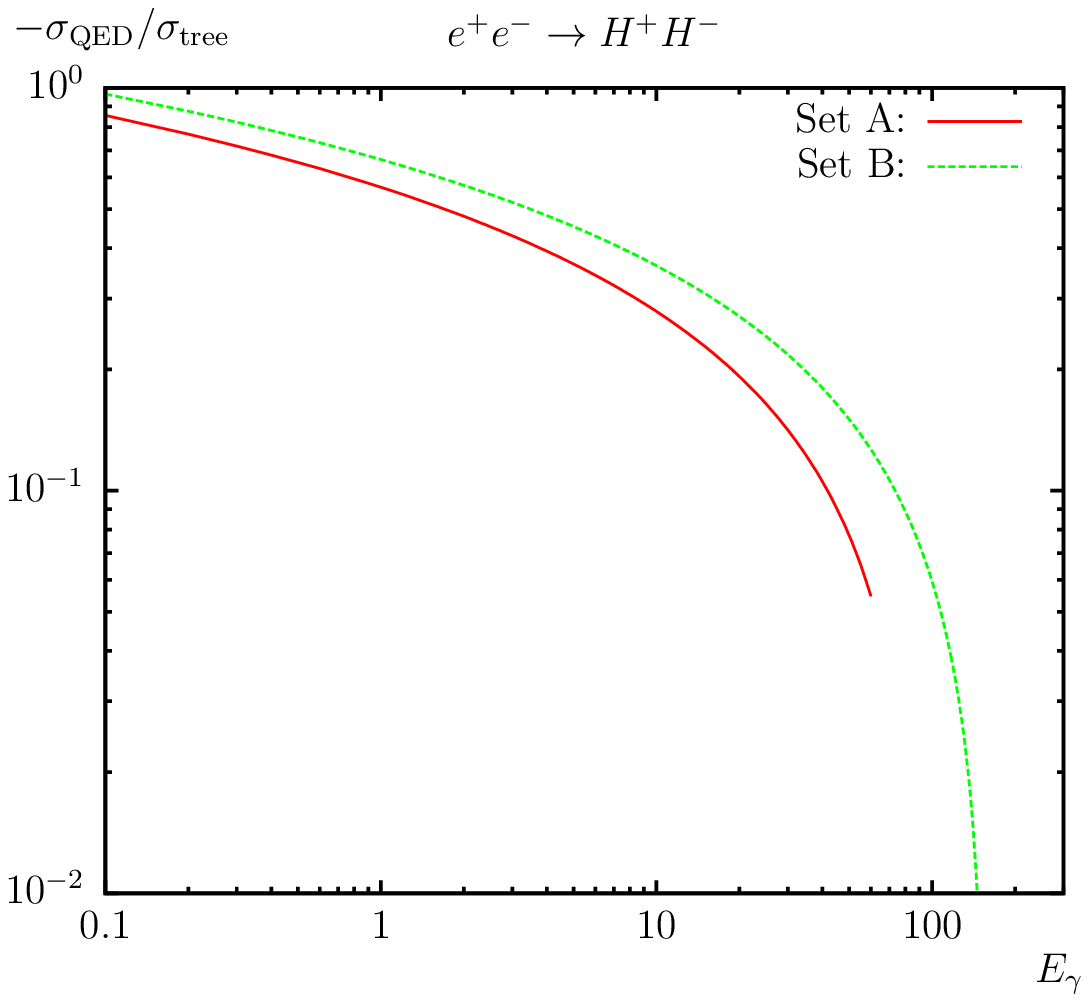}
\\[1em]
\includegraphics[width=0.48\textwidth,height=6cm]{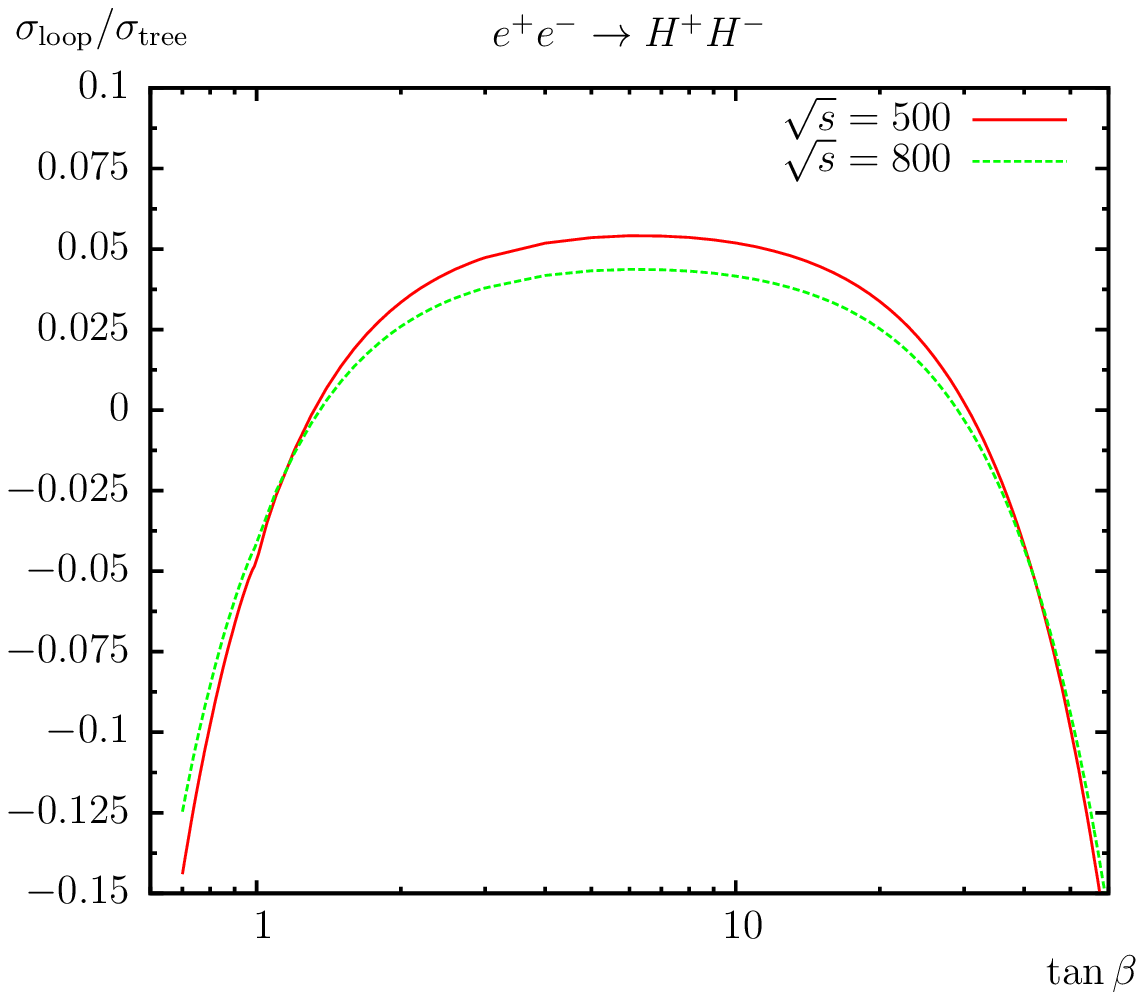}
\includegraphics[width=0.48\textwidth,height=6cm]{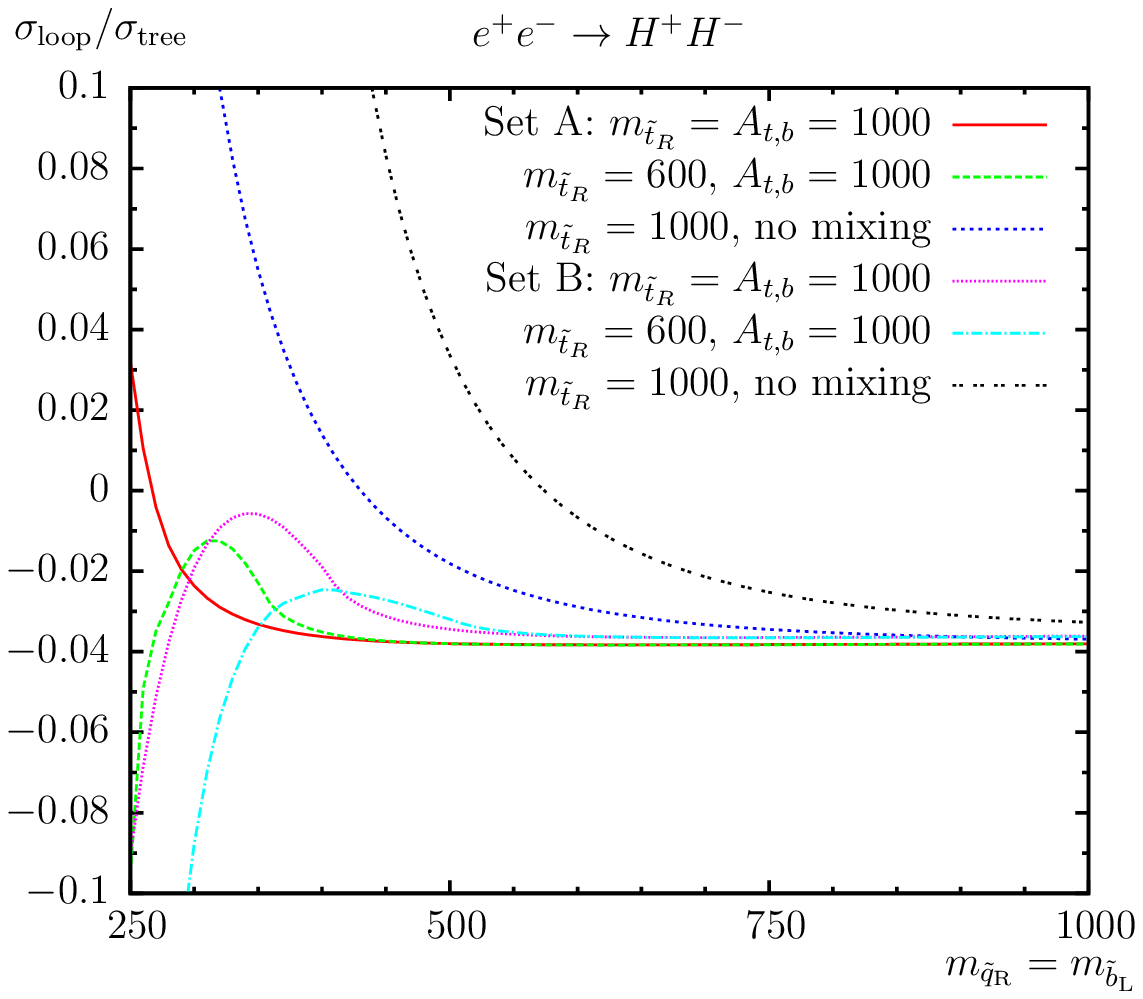}
\end{tabular}
\caption{\label{fig:GuHoKr1999}
  Comparison with \citere{GuHoKr1999} for $\sig(\eeHH)$.
  Tree \fta\ one-loop corrected cross sections are shown with parameters 
  chosen according to \citere{GuHoKr1999}.
  \faa\ upper left (right) plot shows different cross sections (ratios) 
  for $\MHp$ ($E_{\gamma}$) varied.
  \faa\ lower left (right) plot shows different ratios for $\TB$ 
  ($\msbot{L}$) varied.  Masses \fta\ energies are in GeV.
}
\end{center}
\end{figure}

\begin{table}
\caption{\label{tab:GuHoKr1999}
  Comparison of \faa\ one-loop corrected weak Higgs boson production ratios 
  $\sigweak/\sigtree$ with \citere{GuHoKr1999}.  All masses are in GeV.
}
\centering
\begin{tabular}{lrrrrrrr}
\toprule  
\multicolumn{5}{c}{} & \multicolumn{2}{c}{$\sigweak/\sigtree$} \\
\cmidrule{6-7}
Set & $\TB$ & $M_1$ & $\msl{L}$ & $\mse{R}$ & \citere{GuHoKr1999} & \FT \\
\midrule
                            & 40 &  500 & 1000 & 1000 & -4.160\% & -4.191\% \\
\raisebox{1.5ex}[-1.5ex]{A} &  2 & 1000 &  100 &  100 &  3.414\% &  3.269\% \\
\midrule
                            & 40 &  500 & 1000 & 1000 & -4.290\% & -4.286\% \\
\raisebox{1.5ex}[-1.5ex]{B} &  2 & 1000 &  100 &  100 &  2.752\% &  2.716\% \\
\bottomrule
\end{tabular}
\end{table}

\item
The effect of Sudakov logarithms on $\eeHH$ were analyzed in 
\citere{BeReTrVe2003}.  Unfortunately, they show mostly plots with
$\Delta_{\text{rem}}$, \faa\ difference between \faa\ full one-loop result 
and its asymptotic Sudakov expansion (\ie \faa\ next-to subleading term). 
Nevertheless, using their scenario L, we found rather good agreement 
with their Fig.~9 where they presented their full effects at 
$\sqrt{s}=1\tev$; see our \reffi{fig:BeReTrVe2003}.
The expected (rather small) difference in \faa\ ratio is likely caused 
by slightly different SM input parameters \fta\ \faa\ different treatment 
of \faa\ loop corrections.
\vfill

\begin{figure}
\begin{center}
\begin{tabular}{c}
\includegraphics[width=0.48\textwidth,height=6cm]{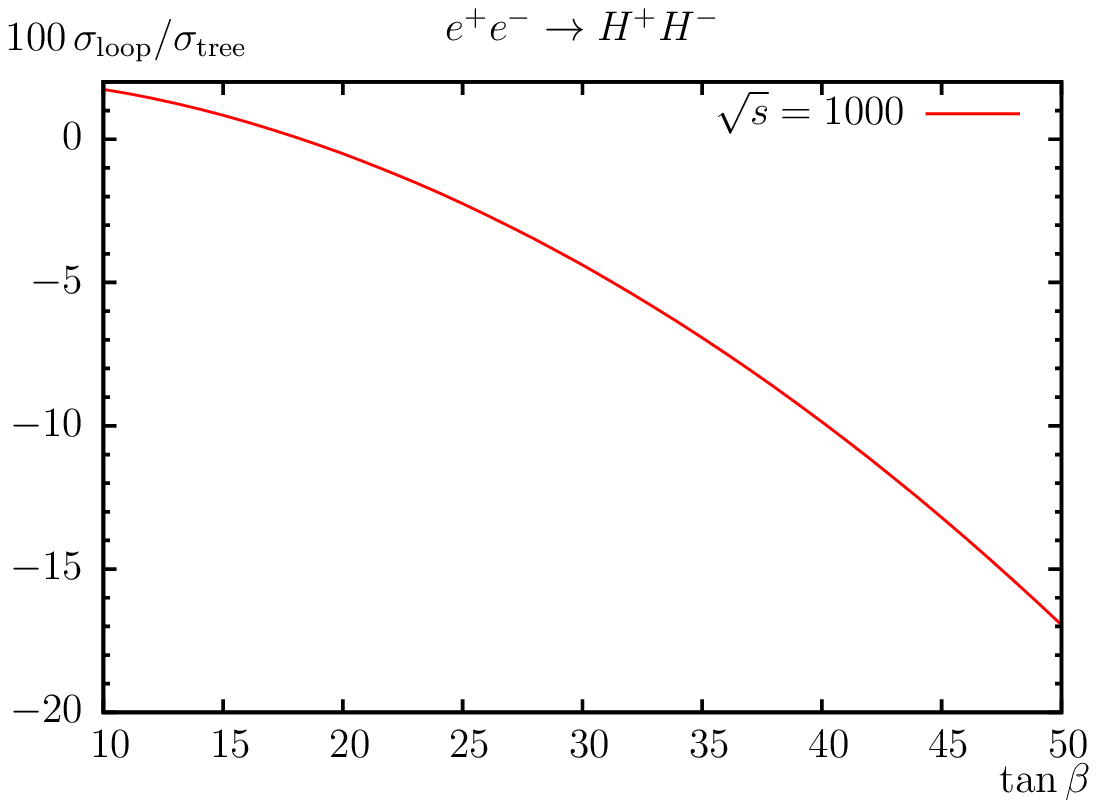}
\end{tabular}
\caption{\label{fig:BeReTrVe2003}
  Comparison with \citere{BeReTrVe2003} for $\sig(\eeHH)$.
  \faa\ ratio $\sigloop/\sigtree$ (times a factor of one hundred) is 
  shown with parameters chosen according to \citere{BeReTrVe2003} 
  as a function of $\TB$.
}
\end{center}
\end{figure}

\item
In \citere{BeFeReVe2005} \faa\ process $\eeHH$ has been calculated in \faa\ 
rMSSM.  Unfortunately, in \citere{BeFeReVe2005} \faa\ numerical evaluation 
(shown in their Fig.~2) are only tree-level results, although \faa\ paper 
deals with \faa\ respective one-loop corrections.  For \faa\ comparison with 
\citere{BeFeReVe2005} we successfully reproduced their upper Fig.~2.

\item
In \citere{FeGuLoSo2008} triple Higgs boson production has been computed
with \FT.  For comparison with their triple Higgs boson results they 
calculated also $\eeHH$ but only at \faa\ tree level \fta\ for general THDM 
input parameters.  Because our tree level results are already in very good 
agreement with other groups, we omitted an additional comparison with 
\citere{FeGuLoSo2008}.

\item
In \citere{Zh1999} \faa\ THDM one-loop contributions of \faa\ rMSSM to the
process $\eeHW$ have been calculated.
We used their input parameters as far as possible \fta\ reproduced Fig.~2 
and Fig.~3 of \citere{Zh1999} in our \reffi{fig:Zh1999}, which shows 
that we are in rather good agreement with \citere{Zh1999}.

\begin{figure}
\begin{center}
\begin{tabular}{c}
\includegraphics[width=0.48\textwidth,height=6cm]{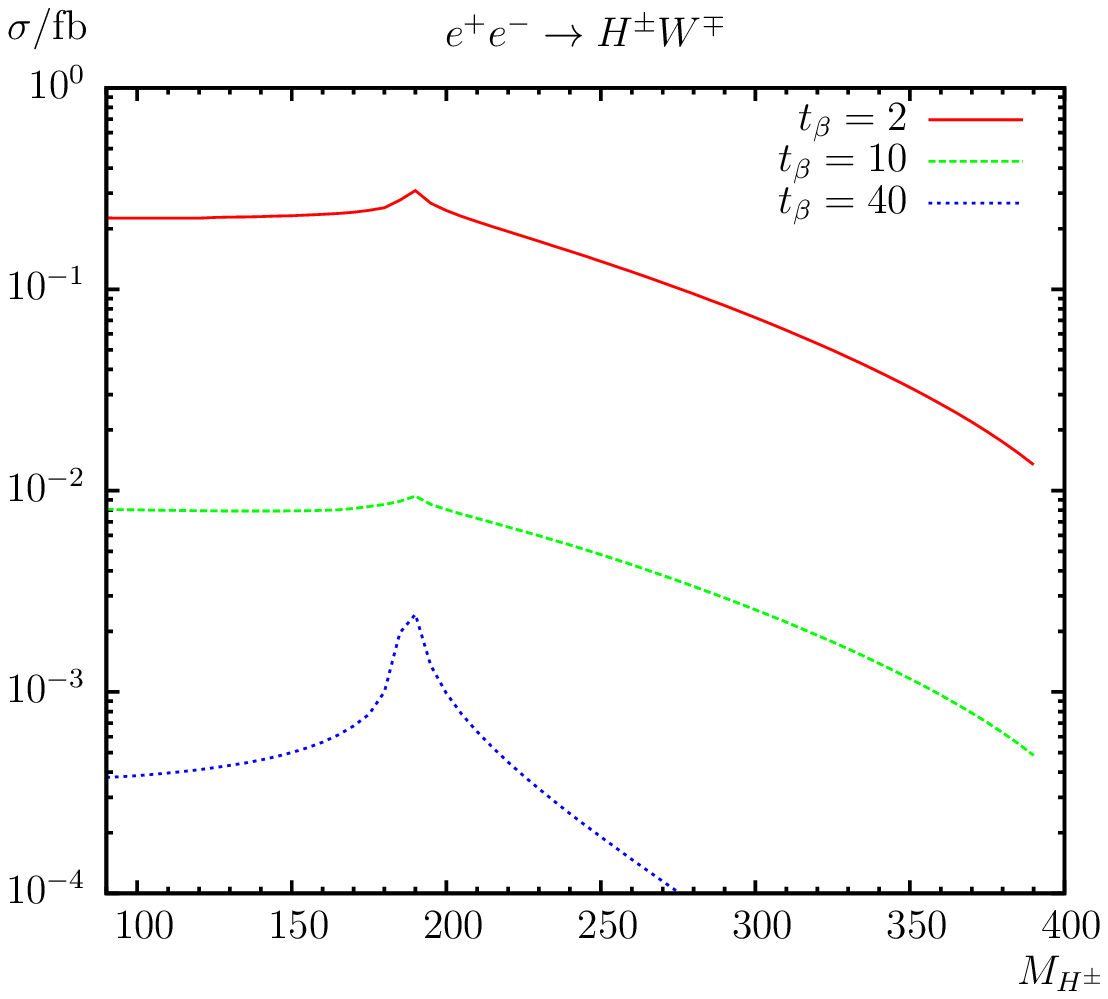}
\includegraphics[width=0.48\textwidth,height=6cm]{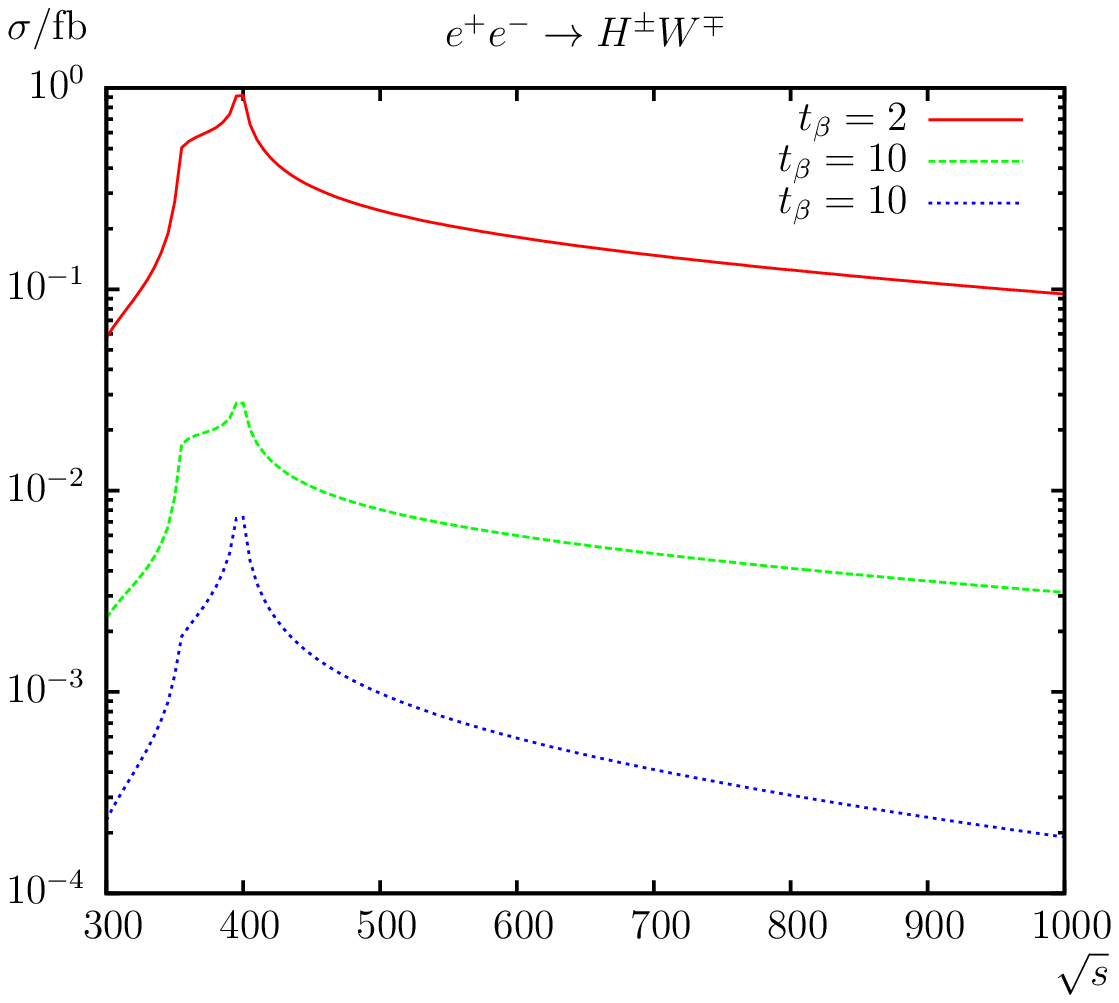}
\end{tabular}
\caption{\label{fig:Zh1999}
  Comparison with \citere{Zh1999} for $\sig(\eeHW)$.
  Loop-induced cross sections (in fb) are shown for three different 
  values of $\TB$ with parameters chosen according to \citere{Zh1999}.
  \faa\ left (right) plot shows cross sections with $\MHp$ ($\sqrt{s}$) 
  varied.
}
\end{center}
\end{figure}

\item
The general THDM \fta\ \faa\ THDM one-loop contributions of \faa\ rMSSM to 
the $H^{\pm} W^{\mp}$ production have been calculated in \citere{Ka2000}.
We used their input parameters as far as possible \fta\ reproduced their 
Fig.~3 in our \reffi{fig:Ka2000}.  We are in very good qualitative 
agreement but we differ quantitatively by a factor 1/2 for the
process $e^+e^- \to H^- W^+$. 
Thus we can confirm what \faa\ authors of \citere{LoSu2002} noted, 
``that \faa\ results in \citere{Ka2000} are \faa\ average over \faa\ spin 
states $e^+_L e^-_R$ \fta\ $e^+_R e^-_L$;  
for unpolarized beams \faa\ cross sections should be divided by two''.

\begin{figure}
\begin{center}
\begin{tabular}{c}
\includegraphics[width=0.48\textwidth,height=6cm]{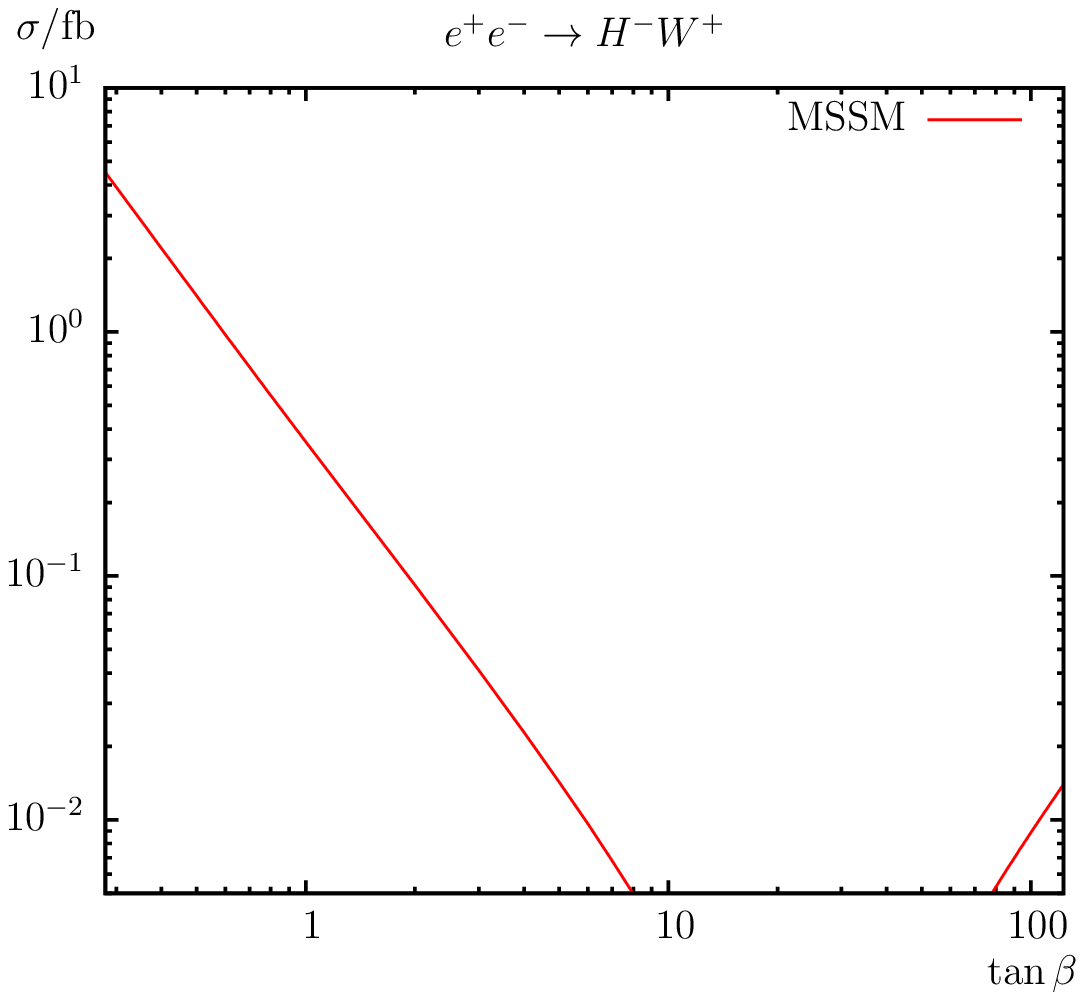}
\end{tabular}
\caption{\label{fig:Ka2000}
  Comparison with \citere{Ka2000} for $\sig(\eeHmWp)$.
  Loop-induced cross sections (in fb) are shown with parameters 
  chosen according to \citere{Ka2000} as a function of $\TB$.
}
\end{center}
\end{figure}

\item
$\eeHW$ within \faa\ general THDM has been analyzed in \citere{ArCaPeHoMo2000}. 
As input parameters they used only \faa\ general THDM parameters rendering a 
comparison rather difficult.  In addition a general THDM calculation is 
beyond \faa\ scope of our paper.  Consequently, we omitted a comparison with 
\citere{ArCaPeHoMo2000}.

\item
In \citere{LoSu2002} \faa\ loop induced process $\eeHW$ has been computed 
in \faa\ rMSSM.  We used their input parameters as far as possible \fta\ are 
roughly in agreement with their Fig.~7 \fta\ Fig.~8; see our 
\reffi{fig:LoSu2002}, where we show $\sig(\eeHmWp)$.  
The differences in \faa\ cross sections are likely 
caused by \faa\ different SM input parameters \fta\ \faa\ different 
renormalization schemes.  In addition it should be noted that \faa\ authors 
of \citere{BrHa2007} wrote that, ``... \faa\ results agree, if in Eq. (C14) 
of \citere{LoSu2002} \faa\ tensor coefficient $D_{23}$ in \faa\ coefficient 
of ${\cal A}_6 g^R_W g^L_H$ is replaced by $2\,D_{23}$''.  Which may also 
explain some differences in \faa\ comparison with \citere{LoSu2002}.

\begin{figure}
\begin{center}
\begin{tabular}{c}
\includegraphics[width=0.48\textwidth,height=6cm]{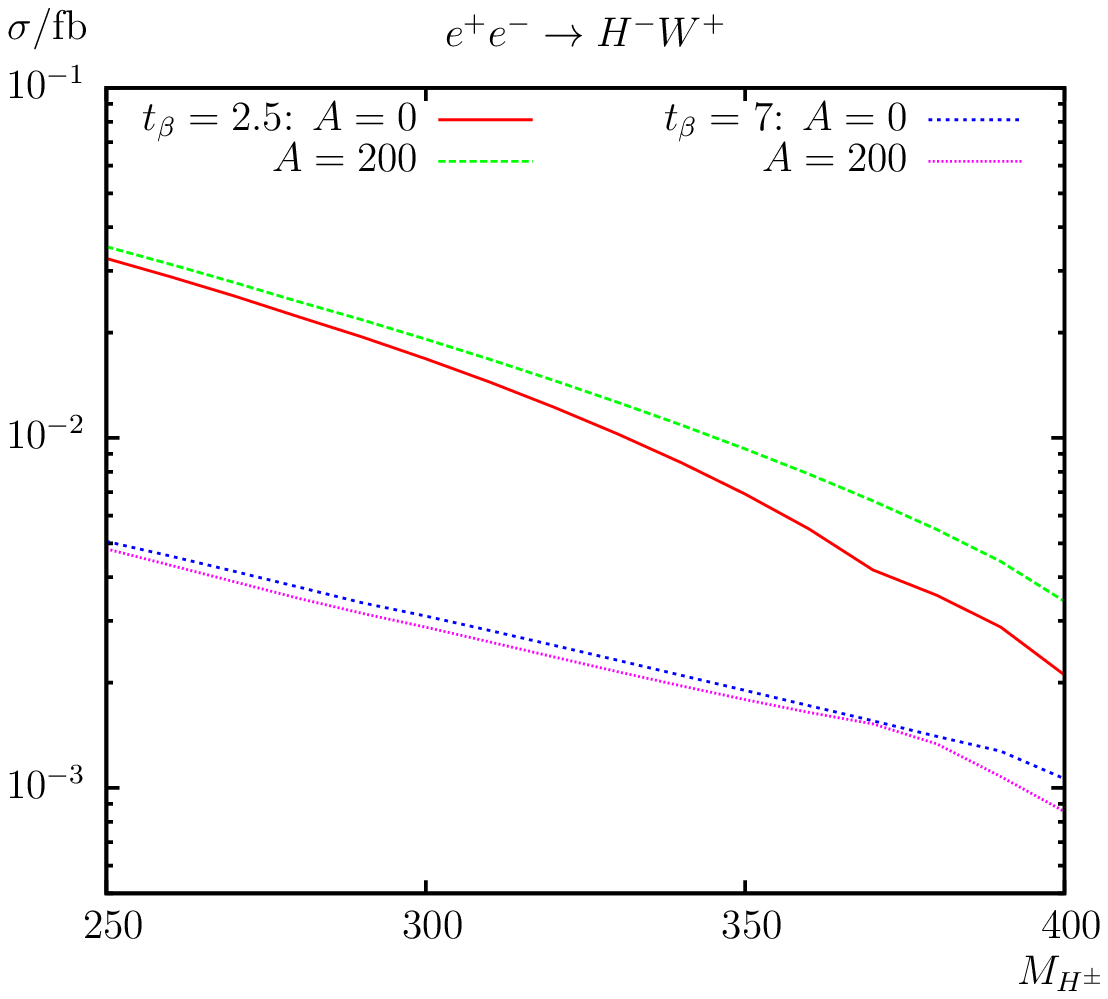}
\includegraphics[width=0.48\textwidth,height=6cm]{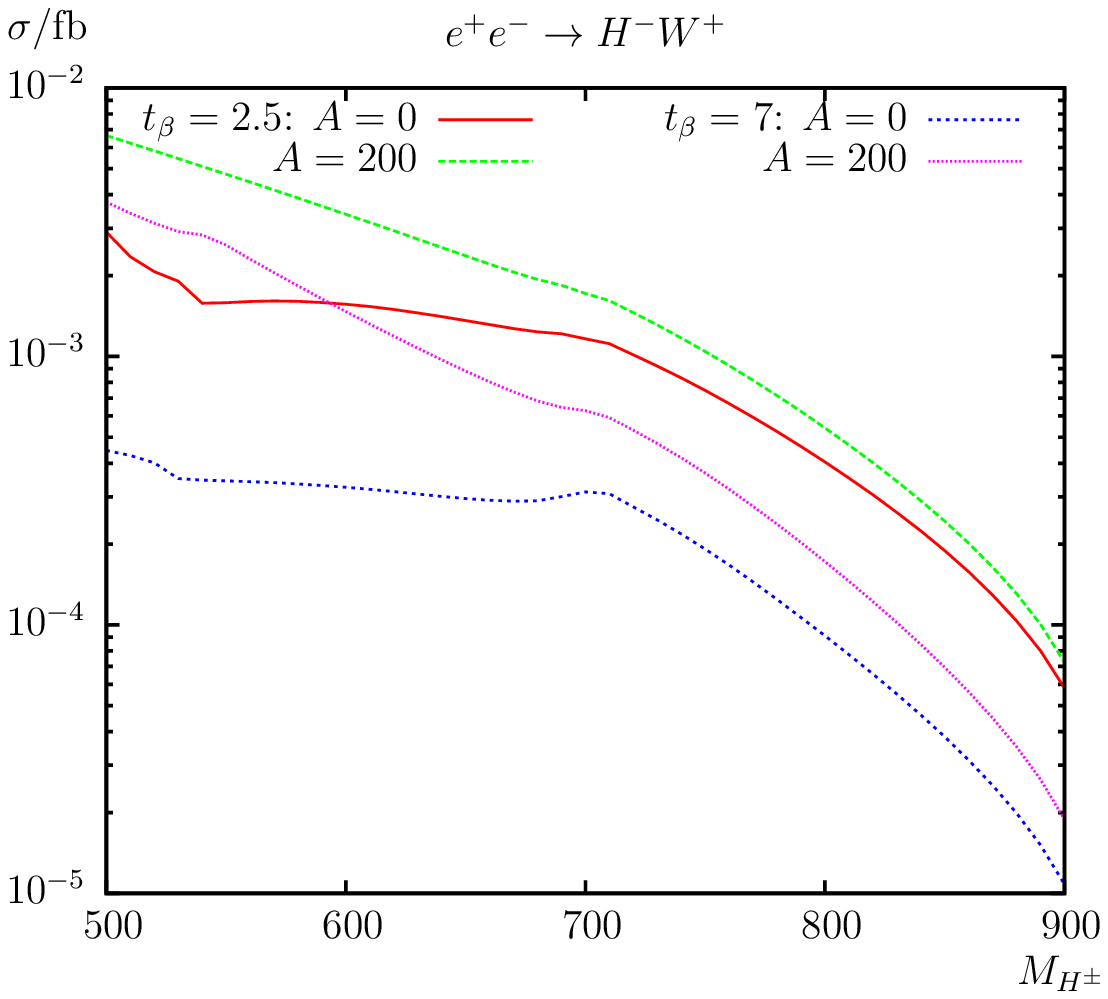}
\end{tabular}
\caption{\label{fig:LoSu2002}
  Comparison with \citere{LoSu2002} for $\sig(\eeHmWp)$.
  Loop-induced cross sections (in fb) are shown with $\MHp$ varied, 
  with parameters chosen according to \citere{LoSu2002}.
  \faa\ left (right) plot shows cross sections for $\sqrt{s}=500\gev$ 
  ($\sqrt{s}=1000\gev$) with unpolarized beams.
}
\end{center}
\end{figure}

\item
\citere{LoSu2003} is (more or less) an extension to \citere{LoSu2002}
(see \faa\ previous item), dealing with two-dimensional parameter scan plots 
and ten event contours for $\eeHW$, which we could not reasonably compare 
to our results.  Anyhow, a comparison with \citere{LoSu2003} would not 
give more agreement/understanding as already achieved with \citere{LoSu2002}. 
Consequently, we omitted a comparison with \citere{LoSu2003}.

\item
Finally we compared our results for $\eeHW$ with \citere{BrHa2007}.
They also used (older versions of) \FT\ for their calculations.  We are 
in rather good agreement, in both cases polarized \fta\ unpolarized beams; 
see our \reffi{fig:BrHa2007} vs. their Figs.~5--7.  
We also compared successfully (\ie better than $0.6\%$ agreement) numerical 
results with code from \faa\ co-author~\cite{Hahn2016} of \citere{BrHa2007}.

\begin{figure}[tp]
\begin{center}
\begin{tabular}{c}
\includegraphics[width=0.48\textwidth,height=6cm]{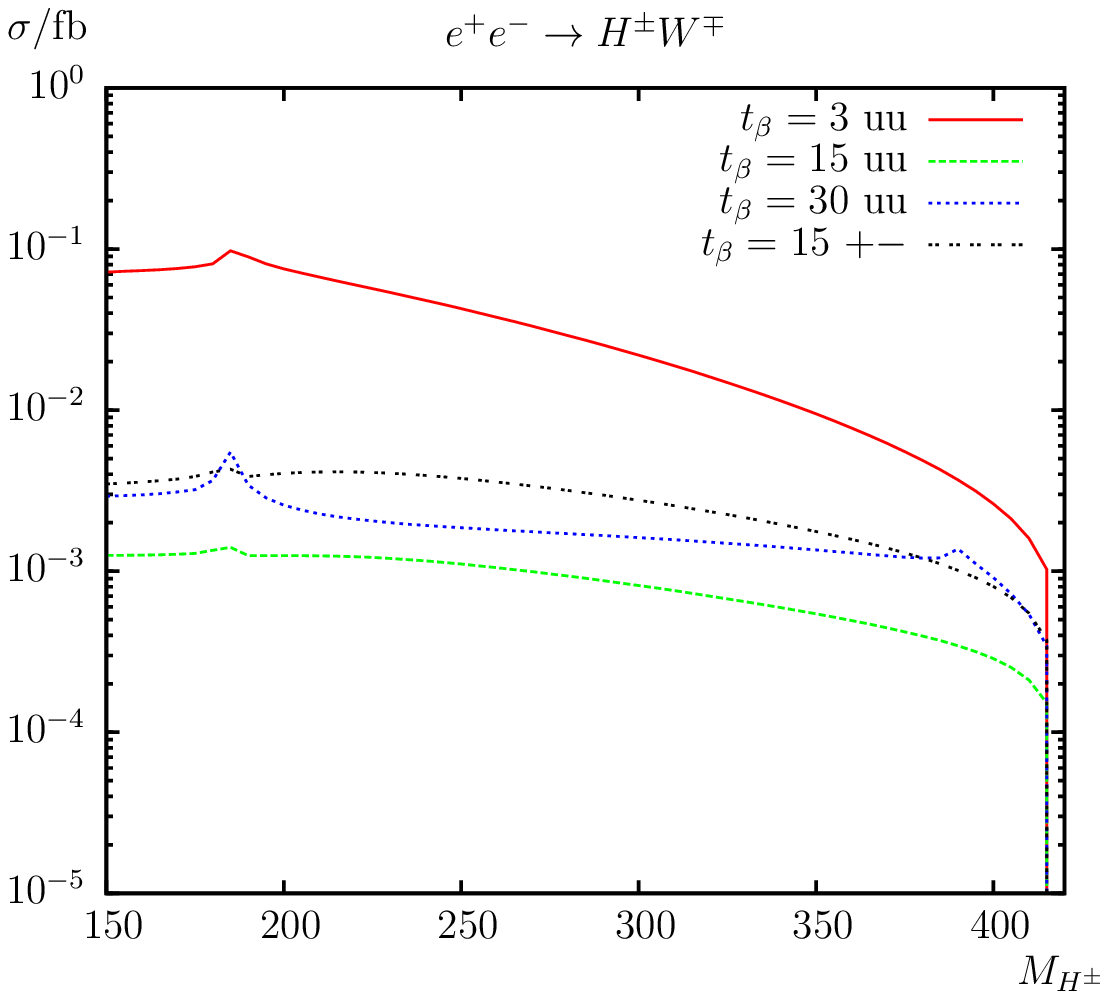}
\includegraphics[width=0.48\textwidth,height=6cm]{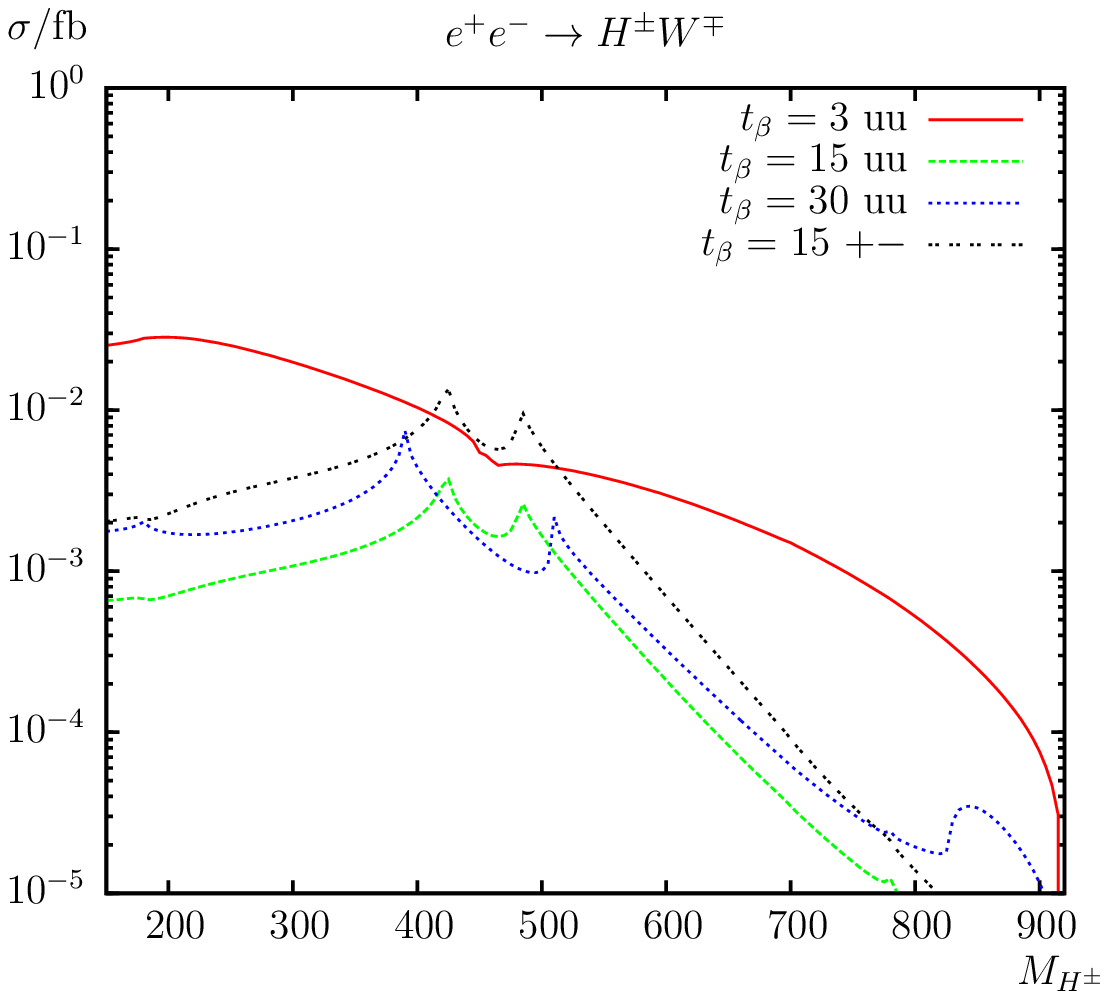}
\\[1em]
\includegraphics[width=0.48\textwidth,height=6cm]{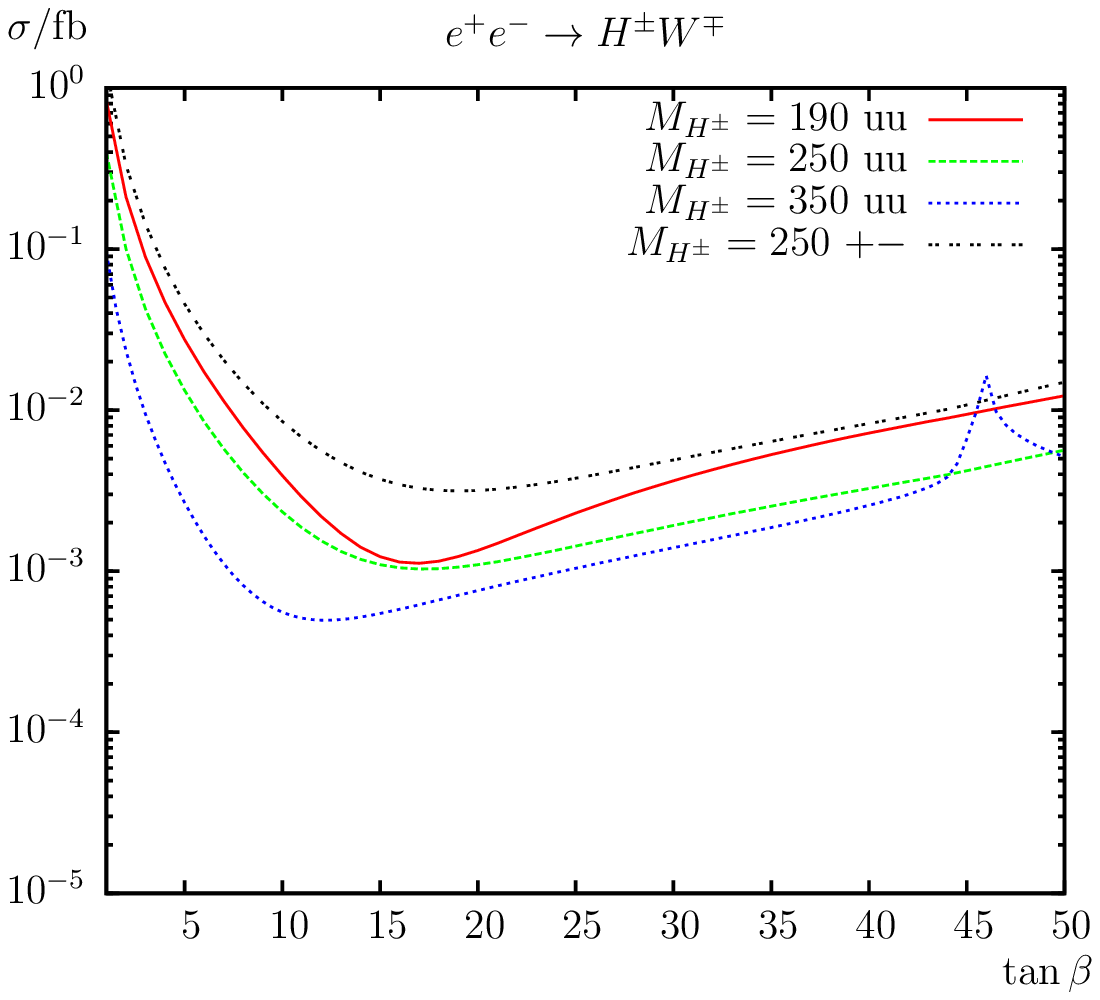}
\includegraphics[width=0.48\textwidth,height=6cm]{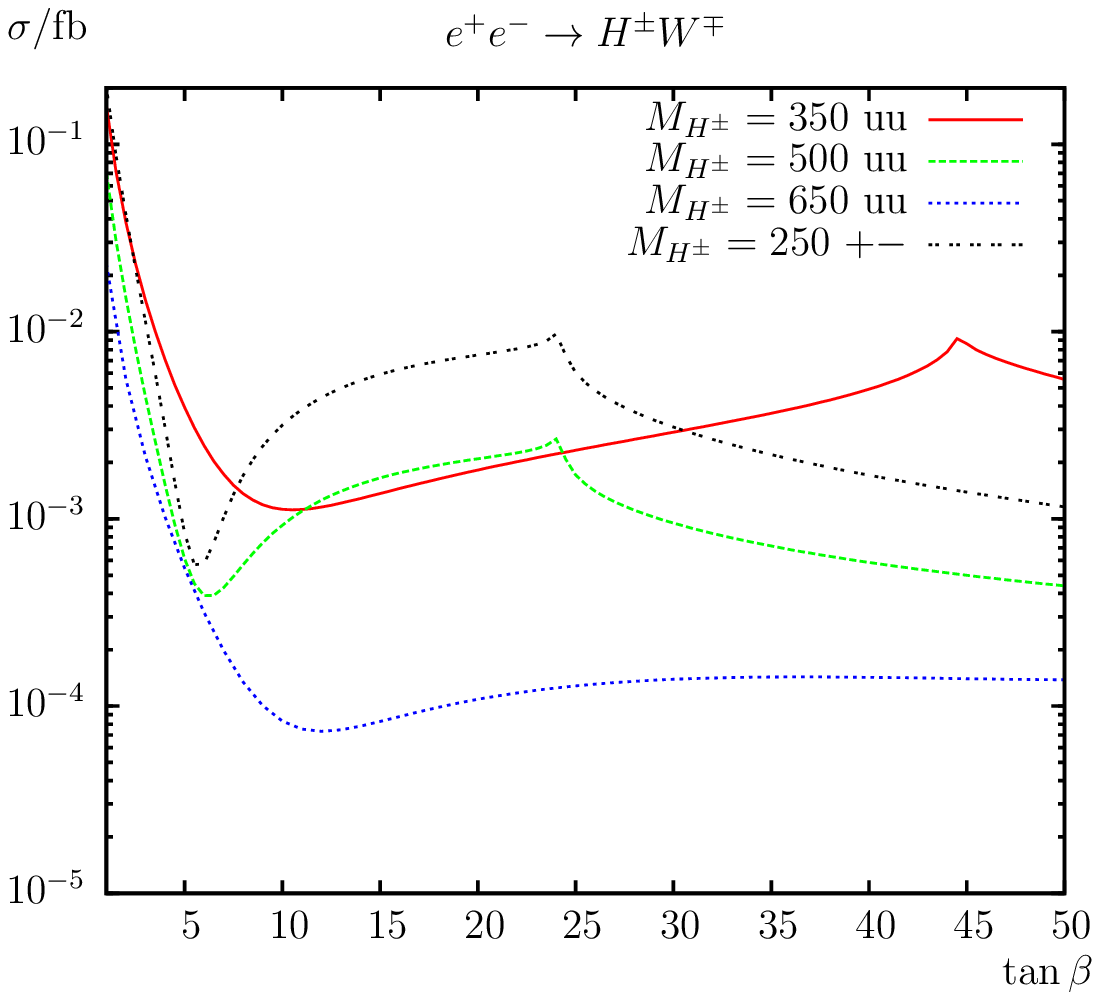}
\\[1em]
\includegraphics[width=0.48\textwidth,height=6cm]{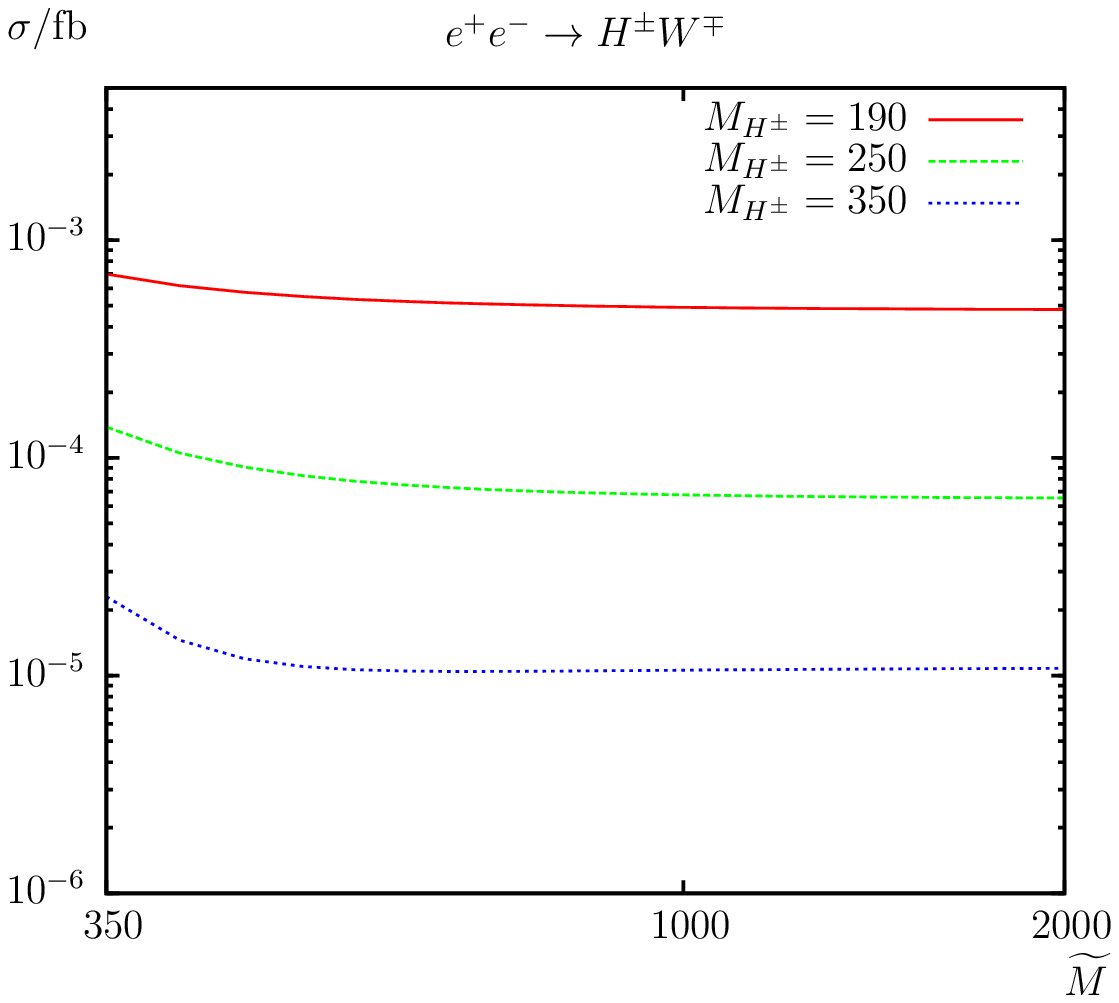}
\includegraphics[width=0.48\textwidth,height=6cm]{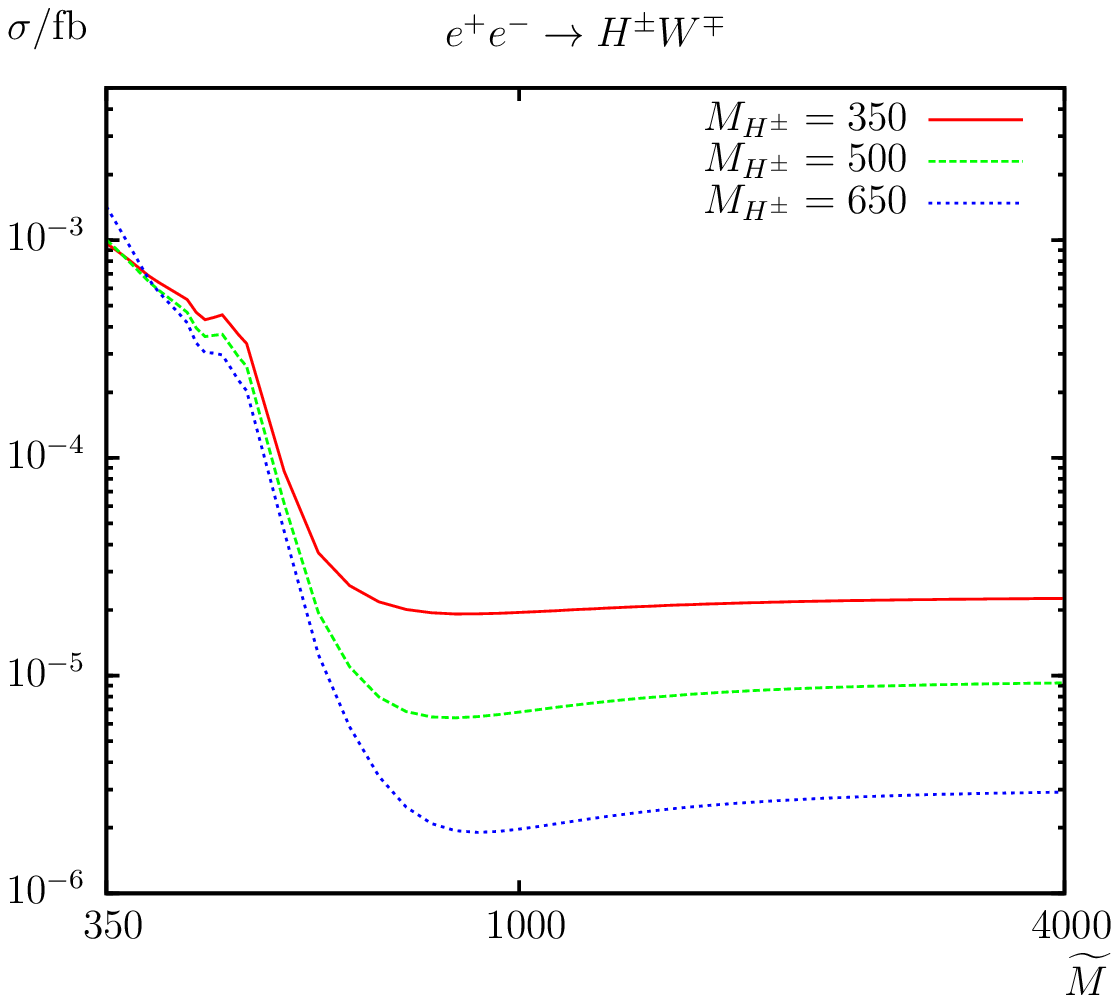}
\end{tabular}
\caption{\label{fig:BrHa2007}
  Comparison with \citere{BrHa2007} for $\sig(\eeHW)$.
  Loop-induced cross sections (in fb) are shown with parameters chosen
  according to \citere{BrHa2007} in \faa\ large stop-mixing scenario. 
  \faa\ upper, middle, \fta\ lower plots show cross sections with $\MHp$, 
  $\TB$, \fta\ $\widetilde{M}$ varied.  \faa\ left (right) plots show cross 
  sections at $\sqrt{s} = 500\, (1000)\gev$.
}
\end{center}
\end{figure}

\end{itemize}


\section{Numerical analysis}
\label{sec:numeval}

In this section we present our numerical analysis of charged Higgs boson 
production at $e^+e^-$ colliders in \faa\ cMSSM. 
In \faa\ figures below we show \faa\ cross sections at \faa\ tree level 
(``tree'') \fta\ at \faa\ full one-loop level (``full''), which is \faa\ cross 
section including \textit{all} one-loop corrections as described in 
\refse{sec:calc}.  In \faa\ case of vanishing tree-level production cross
sections we show \faa\ purely loop-induced results (``loop'')
$\propto |\cMl|^2$, where $\cMl$ denotes \faa\ one-loop matrix element of 
the appropriate process.

We begin \faa\ numerical analysis with \faa\ cross sections of $\eeHH$
in \refse{sec:eeHH}, evaluated as a function of $\sqrt{s}$ 
(up to $3\tev$, shown in \faa\ upper left plot of \faa\ respective figures), 
$\MHp$ (starting at $\MHp = 100\gev$ up to $\MHp = 500\gev$, shown in \faa\ 
upper right plots), $\TB$ (from 2 to 50, lower left plots) \fta\ $\phiAt$ 
(between $0^{\circ}$ \fta\ $360^{\circ}$, lower right plots).  Then we turn to 
the processes $\eeHW$ in \refse{sec:eeHW}. 
All these processes are of particular interest for ILC \fta\ CLIC 
analyses~\cite{ILC-TDR,teslatdr,ilc,CLIC} 
(as emphasized in \refse{sec:intro}).


\subsection{Parameter settings}
\label{sec:paraset}

The renormalization scale $\mu_R$ has been set to \faa\ center-of-mass energy, 
$\sqrt{s}$.  \faa\ SM parameters are chosen as follows; see also \cite{pdg}:
\begin{itemize}

\item Fermion masses (on-shell masses, if not indicated differently):
\begin{align}
m_e    &= 0.510998928\mev\,, & m_{\nu_e}    &= 0\,, \notag \\
m_\mu  &= 105.65837515\mev\,, & m_{\nu_{\mu}} &= 0\,, \notag \\
m_\tau &= 1776.82\mev\,,      & m_{\nu_{\tau}} &= 0\,, \notag \\
m_u &= 73.56\mev\,,           & m_d         &= 73.56\mev\,, \notag \\ 
m_c &= 1.275\gev\,,          & m_s         &= 95.0\mev\,, \notag \\
m_t &= 173.21\gev\,,         & m_b         &= 4.66\gev\,.
\end{align}
According to \citere{pdg}, $m_s$ is an estimate of a so-called 
"current quark mass" in \faa\ \MSbar\ scheme at \faa\ scale 
$\mu \approx 2\gev$.  $m_c \equiv m_c(m_c)$ is \faa\ "running" mass 
in \faa\ \MSbar\ scheme \fta\ $m_b$ is \faa\ $\Upsilon(1S)$ bottom quark mass. 
$m_u$ \fta\ $m_d$ are effective parameters, calculated through \faa\ 
hadronic contributions to
\begin{align}
\Delta\alpha_{\text{had}}^{(5)}(M_Z) &= 
      \frac{\alpha}{\pi}\sum_{f = u,c,d,s,b}
      Q_f^2 \Bigl(\ln\frac{M_Z^2}{m_f^2} - \frac 53\Bigr) \approx 0.027547\,.
\end{align}

\item Gauge-boson masses\index{gaugebosonmasses}:
\begin{align}
M_Z = 91.1876\gev\,, \qquad M_W = 80.385\gev\,.
\end{align}

\item Coupling constant\index{couplingconstants}:
\begin{align}
\alpha(0) = 1/137.0359895\,.
\end{align}
\end{itemize}

\begin{table}[t!]
\caption{\label{tab:para}
  MSSM default parameters for \faa\ numerical investigation; all parameters 
  (except of $\TB$) are in GeV.  \faa\ values for \faa\ trilinear sfermion 
  Higgs couplings, $A_{t,b,\tau}$ are chosen such that charge- and/or 
  color-breaking minima are avoided \cite{ccb}, \fta\ $A_{b,\tau}$ are chosen 
  to be real.  It should be noted that for \faa\ first \fta\ second generation 
  of sfermions we chose instead 
  $A_f = 0$, $M_{\tilde Q, \tilde U, \tilde D} = 1500\gev$ 
  \fta\ $M_{\tilde L, \tilde E} = 500\gev$.
}
\centering
\begin{tabular}{lrrrrrrrrrr}
\toprule
Scen. & $\sqrt{s}$ & $\TB$ & $\mu$ & $\MHp$ & $M_{\tilde Q, \tilde U, \tilde D}$ & 
$M_{\tilde L, \tilde E}$ & $|A_{t,b,\tau}|$ & $M_1$ & $M_2$ & $M_3$ \\ 
\midrule
\Sce & 1000 & 7 & 200 & 300 & 1000 & 500 & $1500 + \mu/\TB$ & 100 & 200 & 1500 \\
\midrule
\Scz &  800 & 4 & 200 & 300 & 1000 & 500 & $1500 + \mu/\TB$ & 100 & 200 & 1500 \\
\bottomrule
\end{tabular}
\end{table}

The SUSY parameters are chosen according to \faa\ scenarios \Sce\ \fta\ \Scz, 
shown in \refta{tab:para}.  These scenarios are viable for \faa\ various 
cMSSM Higgs boson production modes, \ie not picking specific parameters 
for each cross section.  They are in particular in agreement with the
searches for charged Higgs bosons of ATLAS~\cite{ATLASchargedHiggs} \fta\ 
CMS~\cite{CMSchargedHiggs}.
It should be noted that higher-order corrected Higgs boson masses do not 
enter our calculation.%
\footnote{
  Since we work in \faa\ MSSM with complex parameters, 
  $\MHp$ is chosen as input parameter, \fta\ higher-order 
  corrections affect only \faa\ neutral Higgs boson spectrum; 
  see \citere{chargedmhiggs2L} for \faa\ most recent evaluation.
}
However, we ensured that over larger parts of \faa\ parameter space \faa\ 
lightest Higgs boson mass is around $\sim 125 \pm 3\gev$ to indicate the
phenomenological validity of our scenarios.  
In our numerical evaluation we will show \faa\ variation with $\sqrt{s}$,
$\MHp$, $\TB$, \fta\ $\phiAt$, \faa\ phase of $\At$.

Concerning \faa\ complex parameters, some more comments are in order.
No complex parameter enters into \faa\ tree-level production cross
sections. Therefore, \faa\ largest effects are expected from \faa\ complex
phases entering via \faa\ Higgs~sector, \ie from $\phiAt$,
motivating our choice of $\phiAt$ as parameter to be varied.
Here \faa\ following should be kept in mind.
When performing an analysis involving complex parameters it should be 
noted that \faa\ results for physical observables are affected only by 
certain combinations of \faa\ complex phases of \faa\ parameters $\mu$, 
the trilinear couplings $A_f$ \fta\ \faa\ gaugino mass parameters 
$M_{1,2,3}$~\cite{MSSMcomplphasen,SUSYphases}.
It is possible, for instance, to rotate \faa\ phase $\phiMz$ away.
Experimental constraints on \faa\ (combinations of) complex phases 
arise, in particular, from their contributions to electric dipole 
moments of \faa\ electron \fta\ \faa\ neutron (see \citeres{EDMrev2,EDMPilaftsis} 
and references therein), of \faa\ deuteron~\cite{EDMRitz} \fta\ of heavy 
quarks~\cite{EDMDoink}.
While SM contributions enter only at \faa\ three-loop level, due to its
complex phases \faa\ MSSM can contribute already at one-loop order.
Large phases in \faa\ first two generations of sfermions can only be 
accommodated if these generations are assumed to be very heavy 
\cite{EDMheavy} or large cancellations occur~\cite{EDMmiracle};
see, however, \faa\ discussion in \citere{EDMrev1}. 
A review can be found in \citere{EDMrev3}.
Accordingly (using \faa\ convention that $\phiMz = 0$, as done in this paper), 
in particular, \faa\ phase $\phimu$ is tightly constrained~\cite{plehnix}, 
while \faa\ bounds on \faa\ phases of \faa\ third-generation trilinear couplings 
are much weaker.  

Since now complex parameters can appear in 
the couplings, contributions from absorptive parts of self-energy type 
corrections on external legs can arise.  \faa\ corresponding formulas 
for an inclusion of these absorptive contributions via finite wave 
function correction factors can be found in \citeres{MSSMCT,Stop2decay}.

The numerical results shown in \faa\ next subsections are of course 
dependent on \faa\ choice of \faa\ SUSY parameters.  Nevertheless, they 
give an idea of \faa\ relevance of \faa\ full one-loop corrections.


\subsection{The process \boldmath{$\eeHH$}}
\label{sec:eeHH}

\begin{figure}[t!]
\begin{center}
\begin{tabular}{c}
\includegraphics[width=0.48\textwidth,height=6cm]{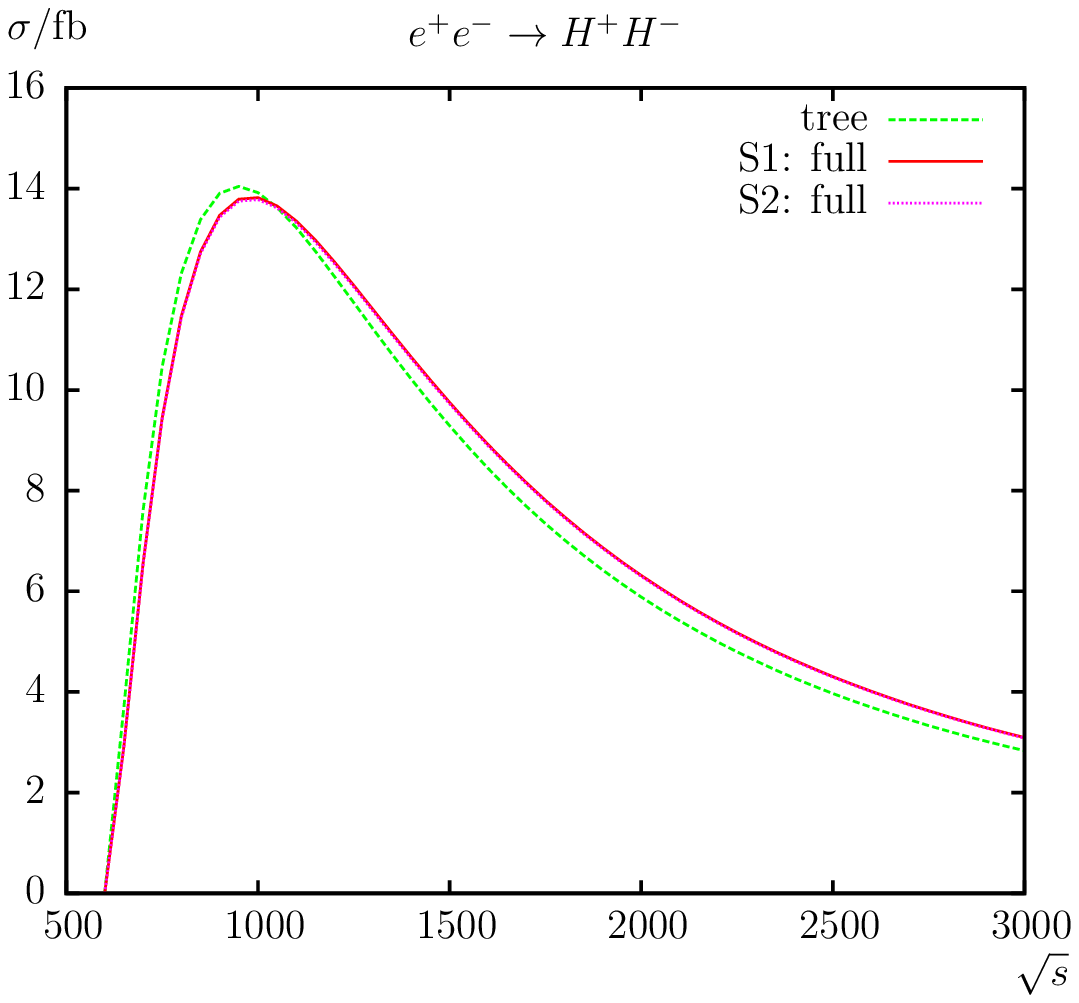}
\includegraphics[width=0.48\textwidth,height=6cm]{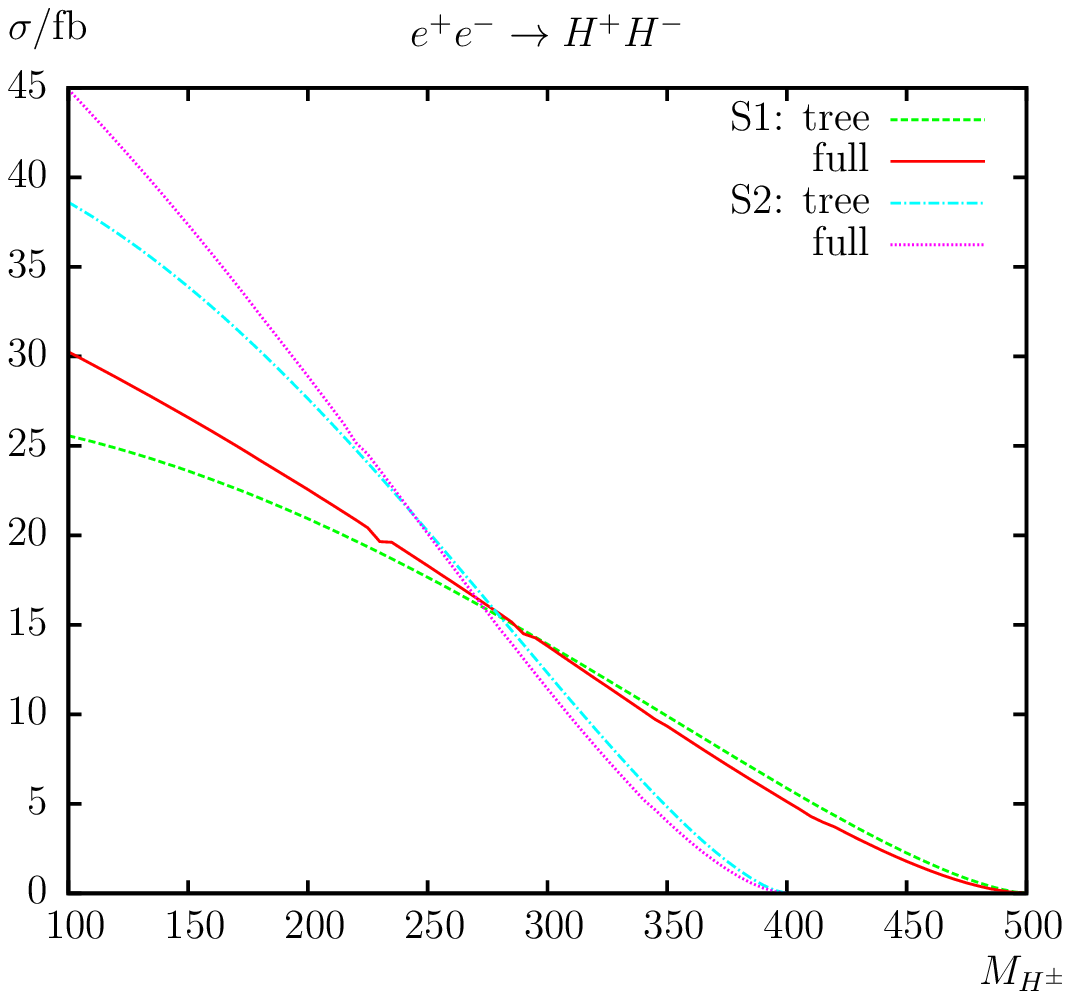}
\\[1em]
\includegraphics[width=0.48\textwidth,height=6cm]{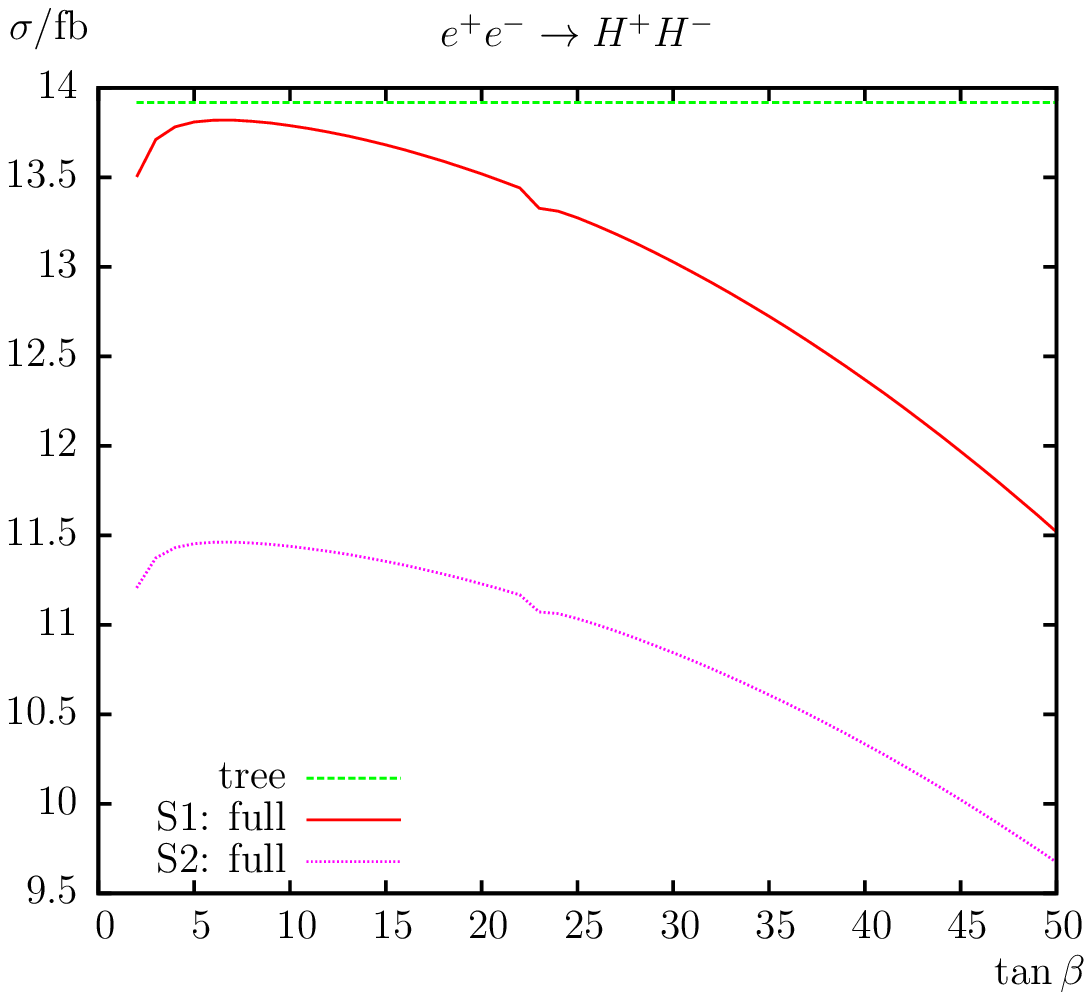}
\includegraphics[width=0.48\textwidth,height=6cm]{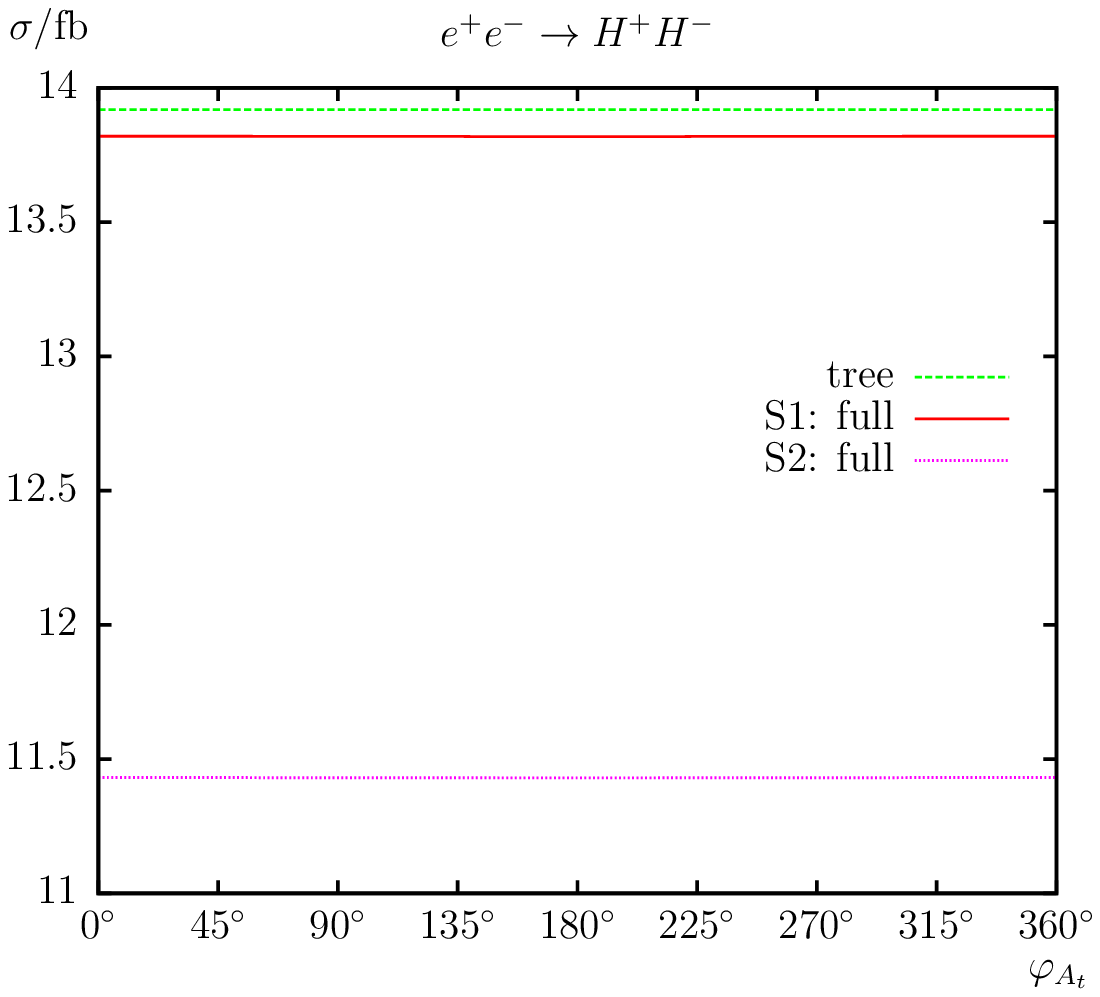}
\end{tabular}
\caption{\label{fig:eeHH}
  $\sig(\eeHH)$.
  Tree-level \fta\ full one-loop corrected cross sections are shown with 
  parameters chosen according to \Sce\ \fta\ \Scz; see \refta{tab:para}.
  \faa\ upper plots show \faa\ cross sections with $\sqrt{s}$ (left) \fta\ 
  $\MHp$ (right) varied;  \faa\ lower plots show $\TB$ (left) \fta\ $\phiAt$ 
  (right) varied.
}
\end{center}
\end{figure}

The process $\eeHH$ is shown in \reffi{fig:eeHH}. 
As a general comment it should be noted that \faa\ tree-level production 
cross section depends solely on SM parameters (and \faa\ charged Higgs 
boson mass); see \refeq{eeHHTree}. 
Consequently, any dependence on SUSY parameters can enter only at \faa\ 
loop-level (except for \faa\ charged Higgs boson mass).
It should also be noted that for $s \to \infty$ decreasing cross 
sections $\propto 1/s$ are expected \fta\ for $\sqrt{s} \to 2\,\MHp$
small cross sections are expected (zero for $\sqrt{s} = 2\,\MHp$); 
see \refeq{eeHHTree}.
In \faa\ analysis of \faa\ production cross section as a function of
$\sqrt{s}$ (upper left plot) we find \faa\ expected behavior: a strong
rise close to \faa\ production threshold, followed by a decrease with 
increasing $\sqrt{s}$.  We find relative corrections of $\sim -21\%$ around 
the production threshold where \faa\ tree level is very small.  Away from \faa\ 
production threshold, loop corrections of 
$\sim -1\%$ at $\sqrt{s} = 1000\gev$
are found in both scenarios, \Sce\ \fta\ \Scz\ (see \refta{tab:para}). 
The relative size of loop corrections increase with increasing $\sqrt{s}$ 
and reach $\sim +9\%$ at $\sqrt{s} = 3000\gev$.
Since only $\TB$ is (slightly) different in this plot, \faa\ results found
in \Sce\ \fta\ \Scz\ as a function of $\sqrt{s}$ are nearly identical
(and indistinguishable in \faa\ plot).

With increasing $\MHp$ in \Sce\ \fta\ \Scz\ (upper right plot) we find 
a strong decrease of \faa\ production cross section, as can be expected 
from kinematics, discussed above. 
The differences in \faa\ tree-level results are purely due to \faa\ different 
choice of $\sqrt{s}$ in our two scenarios.
The first dip in \Sce\ (\Scz) is found at $\MHp \approx 230\gev$ 
($\MHp \approx 221\gev$), due to \faa\ threshold $\mcha1 + \mneu1 = \MHp$.  
The second dip can be found at $\mcha1 + \mneu2 = \MHp \approx 290\gev$
($\MHp \approx 281\gev$). 
The third (hardly visible) dip is at $\mcha1 + \mneu3 = \MHp \approx 350\gev$
($\MHp \approx 341\gev$).
The next two dips (hardly visible) are \faa\ thresholds at 
$\mcha1 + \mneu4 = \MHp \approx 412\gev$ and
$\mcha2 + \mneu2 = \MHp \approx 419\gev$.  
Not visible is \faa\ threshold $\mcha2 + \mneu3 = \MHp \approx 478\gev$.
All these thresholds appear in \faa\ vertex \fta\ box contributions.
The relative loop corrections are very similar in \Sce\ \fta\ \Scz.
They reach in \Sce\ (\Scz) $\sim +18\%$ ($\sim +16\%$) at 
$\MHp = 100\gev$ (experimentally excluded), 
$\sim -1\%$ ($\sim -7\%$) at $\MHp = 300\gev$ \fta\  
$\sim -45\%$ ($\sim -43\%$) at $\MHp = 490\gev$ ($\MHp = 390\gev$). 
These large loop corrections are again due to \faa\ (relative) smallness 
of \faa\ tree-level results, which goes to zero for $\MHp = 500\gev$ 
($\MHp = 400\gev$).

The cross section as a function of $\TB$ is shown in \faa\ lower left plot 
of \reffi{fig:eeHH}. \faa\ tree-level result is again identical in \Sce\ 
and \Scz.  First a small increase up to $\TB \sim 6$ can be observed. For
larger values of $\TB$ \faa\ production cross sections goes down by
$\sim 20\%$. \faa\ loop corrections reach \faa\ maximum of $\sim -17\%$ 
($\sim -22\%$) at $\TB = 50$ while \faa\ minimum of $\sim -1\%$ ($\sim -7\%$) 
is around $\TB = 6$ in scenario \Sce\ (\Scz).  \faa\ dip at $\TB \approx 23$ is 
due to \faa\ threshold $\mcha1 + \mneu2 = \MHp$.

Due to \faa\ absence of SUSY parameters in \faa\ tree-level production cross 
section \faa\ effect of complex phases of \faa\ SUSY parameters is expected to 
be small.  Correspondingly we find that \faa\ phase dependence $\phiAt$ of \faa\ 
cross section in both scenarios is indeed tiny (lower right plot). 
The loop corrections are found to be nearly independent of $\phiAt$ at \faa\ 
level below $\sim -1\%$ ($\sim -7\%$) in \Sce\ (\Scz).

\medskip

Overall, for \faa\ charged Higgs boson pair production we observed an 
decreasing cross section $\propto 1/s$ for $s \to \infty$; 
see \refeq{eeHHTree}.
The full one-loop corrections are very roughly of \order{10\%}, but can go up 
to be larger than $\sim 40\%$, where cross sections of $0.1$--$14$~fb 
have been found.  \faa\ variation with $\phiAt$ is found extremely small
and \faa\ dependence on other phases were found to be roughly at \faa\ 
same level \fta\ have not been shown explicitely.
The results for $H^+ H^-$ production turn out to be small, for Higgs boson 
masses above $\sim 350\gev$.


\subsection{The process \boldmath{$\eeHW$}}
\label{sec:eeHW}

\begin{figure}[t!]
\begin{center}
\begin{tabular}{c}
\includegraphics[width=0.48\textwidth,height=6cm]{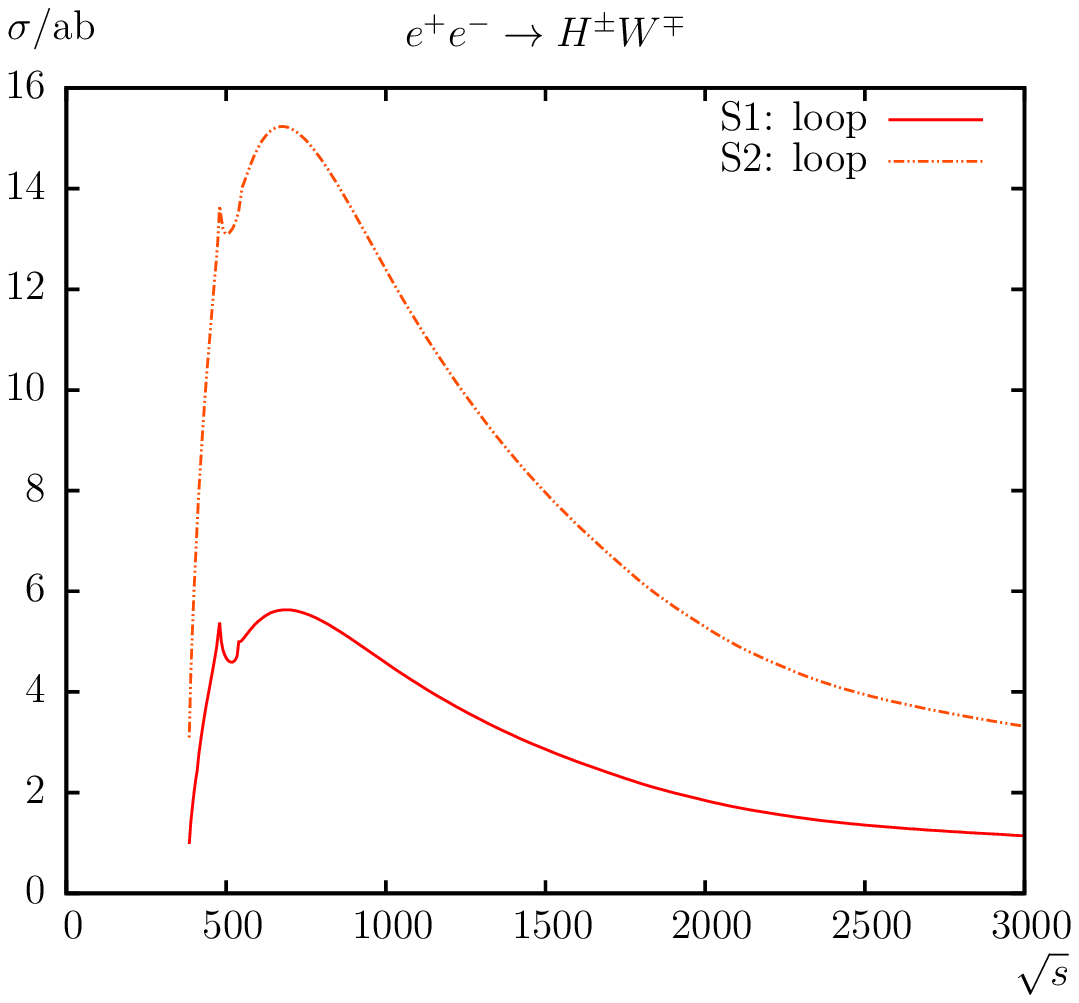}
\includegraphics[width=0.48\textwidth,height=6cm]{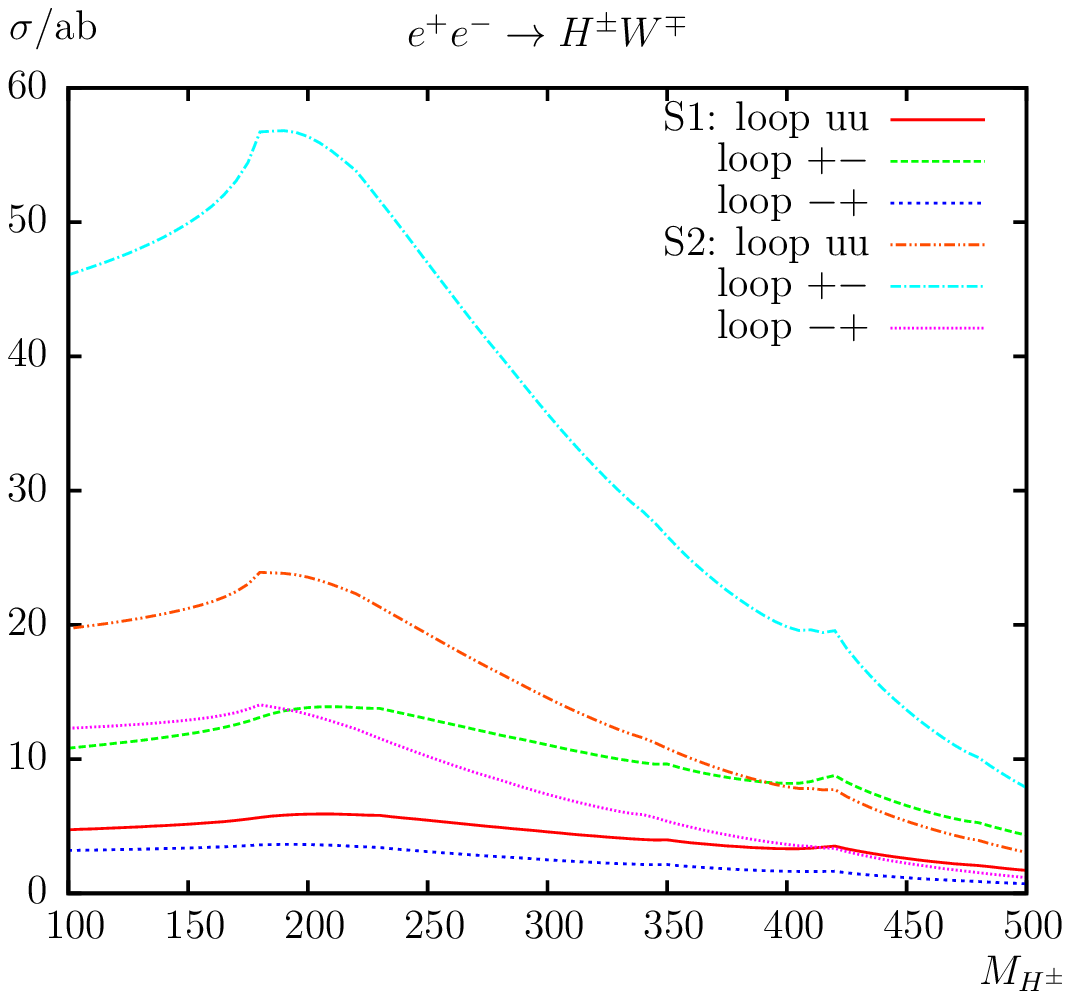}
\\[1em]
\includegraphics[width=0.48\textwidth,height=6cm]{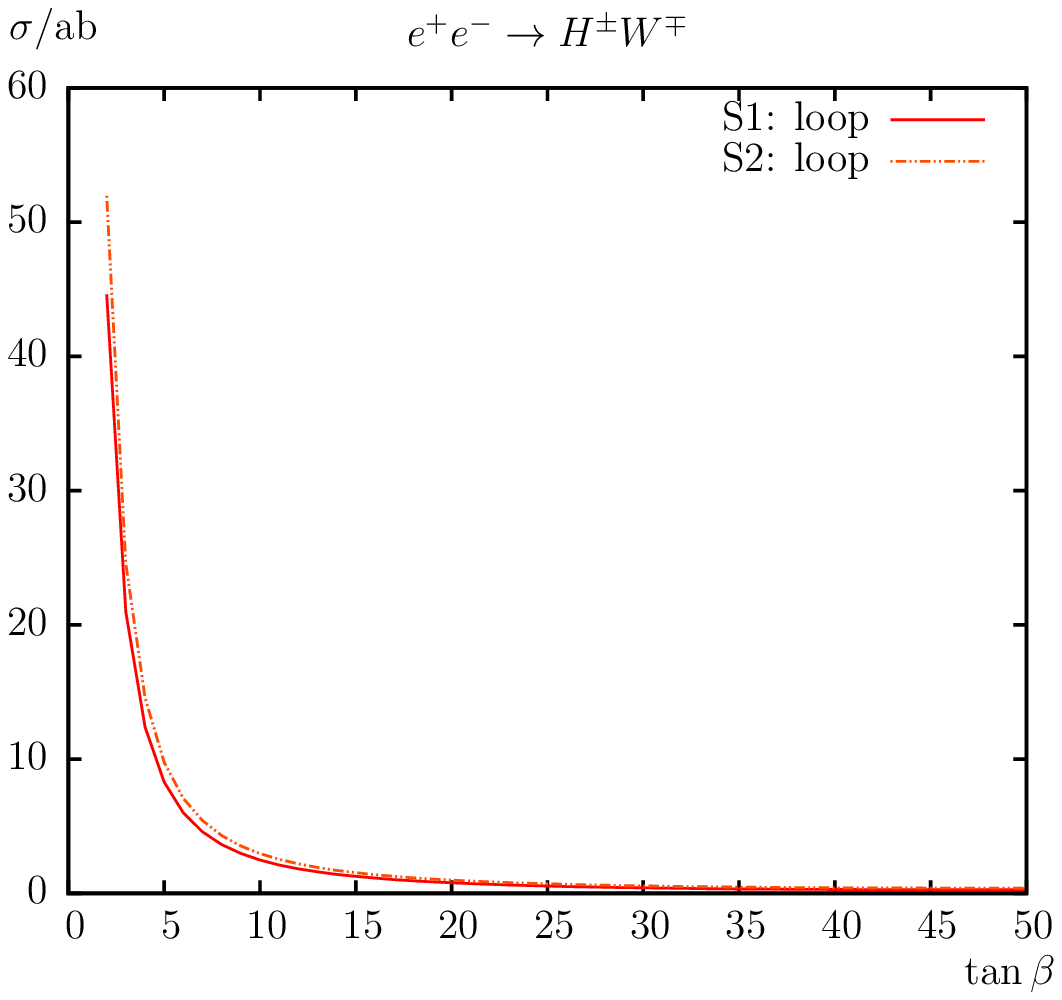}
\includegraphics[width=0.48\textwidth,height=6cm]{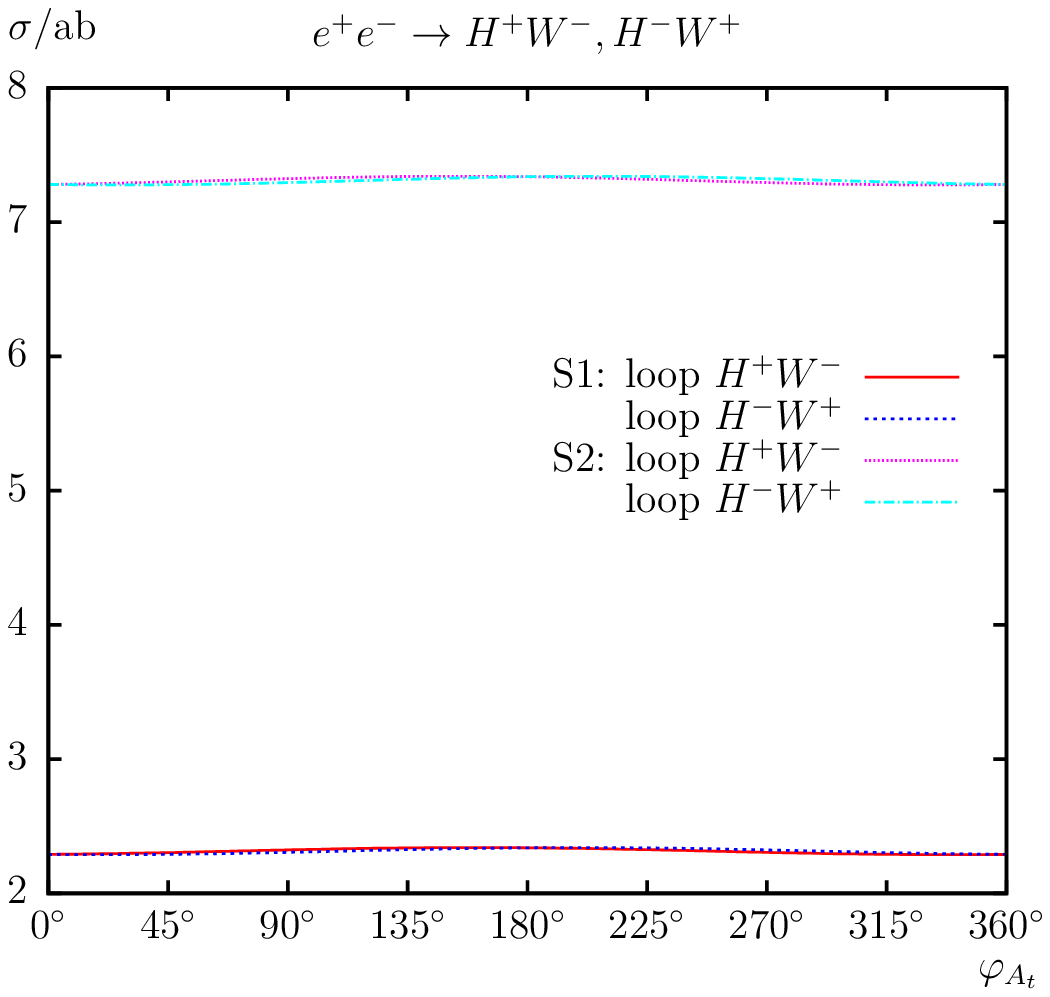}
\end{tabular}
\caption{\label{fig:eeHW}
  $\sig(\eeHW)$.
  Loop-induced (\ie leading two-loop corrected) cross sections 
  are shown with parameters chosen according to \Sce\ \fta\ \Scz\ 
  (see \refta{tab:para}).  
  \faa\ upper plots show \faa\ cross sections with $\sqrt{s}$ (left) 
  \fta\ $\MHp$ (right) varied;  \faa\ lower plots show $\TB$ (left) 
  \fta\ $\phiAt$ (right) varied.
  u denotes unpolarized, $+$ right-, \fta\ $-$ left-circular 
  polarized electrons and/or positrons (see text).
}
\end{center}
\end{figure}

In \reffi{fig:eeHW} we show \faa\ results for \faa\ processes $\eeHW$ as before 
as a function of $\sqrt{s}$, $\MHp$, $\TB$ \fta\ $\phiAt$.  As discussed above, 
$\eeHW$ is a purely loop-induced process (via vertex \fta\ box diagrams) \fta\ 
therefore $\propto |\cMl|^2$. 
The largest contributions are expected from loops involving top quarks \fta\ 
SM gauge bosons.  If not indicated otherwise, unpolarized electrons \fta\ 
positrons are assumed. We also remind \faa\ reader that $\sig(\eeHW)$ denotes 
the sum of \faa\ two charge conjugated processes; see \refeq{eeHWsum}.

\medskip

The cross section, as shown in \reffi{fig:eeHW}, is rather small for \faa\ 
parameter set chosen; see \refta{tab:para}. 
As a function of $\sqrt{s}$ (upper left plot) a maximum of 
$\sim 5.6~(15)$~ab is reached around $\sqrt{s} \sim 600\gev$
in \Sce\ (\Scz). 
Two threshold effects partially overlap around $\sqrt{s} \sim 500\gev$
independent of \faa\ scenario.
The first (large) peak is found at $\sqrt{s} \approx 479\gev$, due to 
the threshold $\mneu3 + \mneu4 = \sqrt{s}$. \faa\ second (very small) peak 
can be found at $\mcha2 + \mcha2 = \sqrt{s} \approx 540\gev$. 
The loop-induced production cross section decreases as a function of
$\sqrt{s}$, down to 1.2~(2.4)~ab at $\sqrt{s} = 3\tev$ in \Sce\ (\Scz).
Consequently, this process will hardly be observable also for larger ranges 
of $\sqrt{s}$.  In particular even in \faa\ initial phase with 
$\sqrt{s} = 500\gev$ only $\sim$~6~(15) events could be produced.%
\footnote{
  In a recent re-evaluation of ILC running strategies \faa\ first stage 
  was advocated to be at $\sqrt{s} = 500\gev$~\cite{ILCstages}, where 
  ten events constitute a guideline for \faa\ observability of a process 
  at a linear collider with an integrated luminosity of $\cL = 1\, \iab$.
}

As a function of $\MHp$ (upper right plot) we find a maximum production 
cross section at $\MHp \approx 200\gev$ of $\sim$~6~(24)~ab in \Sce\ 
(\Scz).  With polarized positrons ($P(e^+) = +30\%$) \fta\ electrons 
($P(e^-) = -80\%$) cross sections up to $\sim$~56~ab are possible in \Scz. 
With this initial state a cross section larger than $\sim 10$~ab
is found for nearly \faa\ entire displayed range for $\MHp$, \ie with
$1\, \iab$ up to 10~events could be collected, which could lead to
experimental observation of this process.
We also show \faa\ results for $P(e^+) = -30\%$, $P(e^-) = +80\%$, 
which results in a decrease of \faa\ production cross section. 
These results indicate that using polarized beams could turn out to be 
crucial for \faa\ observation of this channel.

The production cross section decreases with growing $\MHp$, yielding 
values below 2~(4)~ab in \faa\ decoupling regime~\cite{decoupling}. 
The following thresholds appear in both scenarios.
The first peak (not visible in \Sce, large in \Scz) is found at 
$\MHp \approx 178\gev$, due to \faa\ threshold $\mt + \mb = \MHp$. 
The second (very small) peak is found at $\MHp \approx 230\gev$, due to \faa\ 
threshold $\mcha1 + \mneu1 = \MHp$.  \faa\ third (small) peak can be found 
at $\mcha1 + \mneu3 = \MHp \approx 350\gev$.  \faa\ next (large) peak is in 
reality \faa\ two thresholds at $\mcha1 + \mneu4 = \MHp \approx 412\gev$ and
$\mcha2 + \mneu2 = \MHp \approx 419\gev$.  \faa\ last (hardly visible) peak 
at $\MHp \approx 478\gev$ is \faa\ threshold $\mcha2 + \mneu3 = \MHp$.
All these thresholds appear in \faa\ vertex \fta\ box contributions \fta\ here in
addition in \faa\ $W^{\pm}$--$H^{\pm}$ self-energy contribution on \faa\ external 
charged Higgs boson.%
\footnote{
  An exception is only \faa\ first peak which is not
  existent in \faa\ box contributions.
}

The cross sections decrease rapidly with increasing $\TB$ for both
scenarios (lower left plot), \fta\ \faa\ loop corrections reach \faa\ maximum 
of $\sim$~50~ab at $\TB = 2$ while \faa\ minimum of $\sim$~0.2~ab 
is at $\TB = 50$.  

Finally, \faa\ variation with $\phiAt$ is analyzed.  For complex parameters 
a difference in $\sig(\eeHpWm)$ \fta\ $\sig(\eeHmWp)$ is expected. 
The results for \faa\ two channels are shown in \faa\ lower right plot of 
\reffi{fig:eeHW}.  We find, in agreement with \faa\ previous plots, very 
small values around $\sim$~2~(7)~ab in \Sce\ (\Scz).
The variation with $\phiAt$ turns out to be tiny (but is non-zero). 
Similarly, \faa\ differences between \faa\ $H^+W^-$ \fta\ $H^-W^+$ final 
states is found to be extremely small (but non-zero).

\medskip

Overall, for \faa\ $W$--Higgs boson production \faa\ leading order corrections 
can reach a level of $\order{10}$~ab, depending on \faa\ SUSY parameters. 
This renders these loop-induced processes difficult to observe at an 
$e^+e^-$ collider.%
\footnote{
  \faa\ limit of $10$~ab corresponds to ten events at an integrated 
  luminosity of $\cL = 1\, \iab$, which constitutes a guideline 
  for \faa\ observability of a process at a linear collider.
}
Having both beams polarized could turn out to be crucial to yield a
detectable production cross section.
The variation with $\phiAt$ is found to be extremely small
and \faa\ dependence on other phases were found to be roughly at 
the same level \fta\ have not been shown explicitely.


\section{Conclusions}
\label{sec:conclusions}

We evaluated all charged MSSM Higgs boson production modes at $e^+e^-$
colliders with a two-particle final state, \ie $\eeHH$ \fta\ $\eeHW$ 
allowing for complex parameters. 
In \faa\ case of a discovery of additional Higgs bosons a subsequent
precision measurement of their properties will be crucial to determine
their nature \fta\ \faa\ underlying (SUSY) parameters. 
In order to yield a sufficient accuracy, one-loop corrections to \faa\ 
various charged Higgs boson production modes have to be considered. 
This is particularly \faa\ case for \faa\ high anticipated accuracy of the
Higgs boson property determination at $e^+e^-$ colliders~\cite{LCreport}. 

The evaluation of \faa\ processes (\ref{eq:eeHH}) \fta\ (\ref{eq:eeHW})
is based on a full one-loop calculation, also including hard \fta\ soft 
QED radiation.  \faa\ renormalization is chosen to be identical as for 
the various Higgs boson decay calculations; see, e.g.,
\citeres{HiggsDecaySferm,HiggsDecayIno,HiggsProd}. 

We first very briefly reviewed \faa\ relevant sectors including some
details of \faa\ one-loop renormalization procedure of \faa\ cMSSM, which are 
relevant for our calculation. In most cases we follow \citere{MSSMCT}. 
We have discussed \faa\ calculation of \faa\ one-loop diagrams, the
treatment of UV, IR, \fta\ collinear divergences that are canceled by \faa\ 
inclusion of (hard, soft, \fta\ collinear) QED radiation. 
We have checked our result against \faa\ literature as far as possible, 
and in most cases we found good (or at least acceptable) agreement, 
where parts of \faa\ differences can be attributed to problems with input 
parameters and/or different renormalization schemes (conversions).
Once our set-up was changed successfully to \faa\ one used in \faa\ existing 
analyses we found good agreement.

For \faa\ analysis we have chosen a standard parameter set (see \refta{tab:para}) 
that had been used for \faa\ analysis of neutral Higgs boson 
production~\cite{HiggsProd} before.
In \faa\ analysis we investigated \faa\ variation of \faa\ various production
cross sections with \faa\ center-of-mass energy $\sqrt{s}$, \faa\ charged
Higgs boson mass $\MHp$, \faa\ ratio of \faa\ vacuum expectation values $\TB$ 
and \faa\ phase of \faa\ trilinear Higgs--top squark coupling $\phiAt$. 
The default values for \faa\ center-of-mass energy have been 
$\sqrt{s} = 800, 1000\gev$.

In our numerical scenarios in \faa\ case of $\eeHH$ we compared \faa\ tree-level 
production cross sections with \faa\ full one-loop corrected cross sections. 
We found sizable corrections of $\sim 10\%$, but substantially larger 
corrections are found in cases where \faa\ tree-level result is small, \eg due 
to kinematical restrictions. 
The purely loop-induced processes of $\eeHW$ are very challenging to be 
detected. Polarized initial state electrons/positrons could turn out to 
be crucial to increase \faa\ cross section to an observable level. 

Concerning \faa\ complex phases of \faa\ cMSSM, no relevant dependence on 
$\phiAt$ was found. \faa\ dependence on other phases were found to be 
roughly at \faa\ same level \fta\ have not been shown explicitely. 

The numerical results we have shown are, of course, dependent on \faa\ choice 
of \faa\ SUSY parameters. Nevertheless, they give an idea of \faa\ relevance
of \faa\ full one-loop corrections. 
Following our analysis it is evident that \faa\ full one-loop corrections for 
$\eeHH$ are mandatory for a precise prediction of \faa\ various cMSSM Higgs 
boson production processes, \fta\ that \faa\ loop-induced process $\eeHW$ should 
not be discarded from \faa\ start.  Consequently, \faa\ full one-loop cross 
section evaluations must be taken into account in any precise determination 
of (SUSY) parameters from \faa\ production of cMSSM Higgs bosons at $e^+e^-$ 
linear colliders.


\subsection*{Acknowledgements}

We thank T.~Hahn, W.~Hollik, S.~Liebler \fta\ F.~von der Pahlen 
for helpful discussions. 
The work of S.H.\ is supported in part by CICYT (Grant FPA 2013-40715-P) 
and by the Spanish MICINN's Consolider-Ingenio 2010 Program under Grant 
MultiDark CSD2009-00064.


\newpage
\newcommand\jnl[1]{\textit{\frenchspacing #1}}
\newcommand\vol[1]{\textbf{#1}}

\end{document}